\documentclass{sig-alternate-10pt}

\usepackage[caption=false]{subfig}
\usepackage{balance}

\newcommand{\comments}[1]{}

\newtheorem{theo}{Theorem}

\newtheorem{defin}{Definition}
\newtheorem{lem}{Lemma}
\newtheorem{cor}{Corollary}

\newcommand{\eqdef}{\stackrel{\triangle}{=}}

\begin{document}

\title{XORing Elephants: Novel Erasure Codes for Big Data}

\numberofauthors{7}
\author{
\alignauthor
Maheswaran Sathiamoorthy\\
      \affaddr{University of Southern California}\\
       \email{msathiam@usc.edu}
\alignauthor
Megasthenis Asteris\\
       \affaddr{University of Southern California}\\
       \email{asteris@usc.edu}
\alignauthor
Dimitris Papailiopoulos\\ 
       \affaddr{University of Southern California}\\
       \email{papailio@usc.edu}
\and
\alignauthor Alexandros G. Dimakis\\
       \affaddr{University of Southern California}\\
       \email{dimakis@usc.edu}
\alignauthor
 Ramkumar Vadali\\
       \affaddr{Facebook}\\
       \email{ramkumar.vadali@fb.com}
 \alignauthor
  Scott Chen\\
       \affaddr{Facebook}\\
       \email{sc@fb.com}
\and
 \alignauthor
  Dhruba Borthakur\\
   \affaddr{Facebook}
  \email{dhruba@fb.com}
  }

\maketitle
  
\begin{abstract}
Distributed storage systems for large clusters typically use replication to provide reliability. Recently, erasure codes have been used to reduce the large storage overhead of three-replicated systems. Reed-Solomon codes are the standard design choice and their high repair cost is often considered an unavoidable price to pay for high storage efficiency and high reliability.

This paper shows how to overcome this limitation. We present a novel family of erasure codes that are efficiently repairable and offer higher reliability compared to Reed-Solomon codes. We show analytically that our codes are optimal on a recently identified tradeoff between locality and minimum distance.

We implement our new codes in Hadoop HDFS and compare to a currently deployed HDFS module that uses Reed-Solomon codes. Our modified HDFS implementation shows a reduction of approximately $2\times$ on the repair disk I/O and repair network traffic. The disadvantage of the new coding scheme is that it requires 14\% more storage compared to Reed-Solomon codes, an overhead shown to be information theoretically optimal to obtain locality.
Because the new codes repair failures faster, this provides higher
reliability, which is orders of magnitude higher compared to replication.
\end{abstract}

\section{Introduction}
\label{sec:intro}

MapReduce architectures are becoming increasingly popular for big data management 
due to their high scalability properties. At Facebook, large analytics clusters store petabytes of information and handle multiple analytics jobs using Hadoop MapReduce. Standard implementations rely on a distributed file system that provides reliability by exploiting triple block replication. The major disadvantage of replication is the very large storage overhead of $200\%$, which reflects on the cluster costs. This overhead is becoming a major bottleneck 
as the amount of managed data grows faster than data center infrastructure. 

For this reason, Facebook and many others are transitioning to erasure coding techniques (typically, classical Reed-Solomon codes) to introduce redundancy while saving storage
~\cite{Azure_Storage,Khan_Fast12}, especially for data that is more archival in nature. In this paper we show that classical codes are highly suboptimal for distributed MapReduce architectures. We introduce new erasure codes that address the main challenges of distributed data reliability and information theoretic bounds that show the optimality of our construction. We rely on measurements from a large Facebook production cluster (more than $3000$ nodes, $30$ PB of logical data storage) that uses Hadoop MapReduce for data analytics.
Facebook recently started deploying an open source HDFS Module called HDFS RAID (\cite{hdfsraidwiki,diskreduce}) that relies on Reed-Solomon (RS) codes. 
In HDFS RAID, the replication factor of ``cold'' (\textit{i.e.}, rarely accessed) files is lowered to $1$ and a new parity file is created, consisting of parity blocks. 

Using the parameters of Facebook clusters, the data blocks of each large file are grouped in stripes of $10$ and for each such set, $4$ parity blocks are created. This system (called RS $(10,4)$) can tolerate any $4$ block failures and has a storage overhead of only $40\%$. RS codes are therefore significantly more robust and storage efficient compared to replication. In fact, this storage overhead is the minimal possible, for this level of reliability~\cite{dimakis2011survey}.
Codes that achieve this optimal storage-reliability tradeoff are called Maximum Distance Separable (MDS)
\cite{Wicker} and Reed-Solomon codes~\cite{ReedSolomon} form the most widely used MDS family. 

Classical erasure codes are suboptimal for distributed environments because of the 
so-called~\textit{Repair problem:} When a single node fails, typically one block is lost from each stripe that is stored in that node. RS codes are usually repaired with the simple method that requires transferring $10$ blocks and recreating the original $10$ data blocks even if a single block is lost~\cite{rodrigues2005high}, hence creating a $10 \times$ overhead in repair bandwidth and disk I/O. 

Recently, information theoretic results established that it is possible to repair erasure codes with much less network bandwidth compared to this naive method~\cite{dimakis2010network}.
There has been significant amount of very recent work on designing such efficiently repairable codes, see section~\ref{sec:relatedwork} for an overview of this literature.
 
\textbf{Our Contributions:} We introduce a new family of erasure codes called \textit{Locally Repairable Codes (LRCs)}, that are efficiently repairable both in terms of network bandwidth and disk I/O. We analytically show that our codes are information theoretically optimal in terms of their locality, \textit{i.e.}, the number of other blocks needed to repair \emph{single block failures}. We present both randomized and explicit LRC constructions starting from generalized Reed-Solomon parities. 

We also design and implement \textit{HDFS-Xorbas}, a module that replaces Reed-Solomon codes with LRCs in HDFS-RAID. We evaluate HDFS-Xorbas using experiments on Amazon EC2 and a cluster in Facebook. Note that while LRCs are defined for any stripe and parity size, our experimental evaluation is based on a RS(10,4) and its extension to a (10,6,5) LRC to compare with the current production cluster. 

Our experiments show that Xorbas enables approximately a $2\times$ reduction in disk I/O and repair network traffic compared to the Reed-Solomon code currently used in production. 
The disadvantage of the new code is that it requires $14\%$ more storage compared to RS, an overhead shown to be information theoretically optimal for the obtained locality.

One interesting side benefit is that because Xorbas repairs failures faster, this provides higher availability, due to more efficient degraded reading performance. Under a simple Markov model evaluation, Xorbas has 2 more zeros in Mean Time to Data Loss (MTTDL) compared to RS $(10,4)$ and 5 more zeros compared to 3-replication. 

\subsection{Importance of Repair}

\begin{figure}[t!]
	\hspace{0cm}\includegraphics[width=0.45\textwidth]{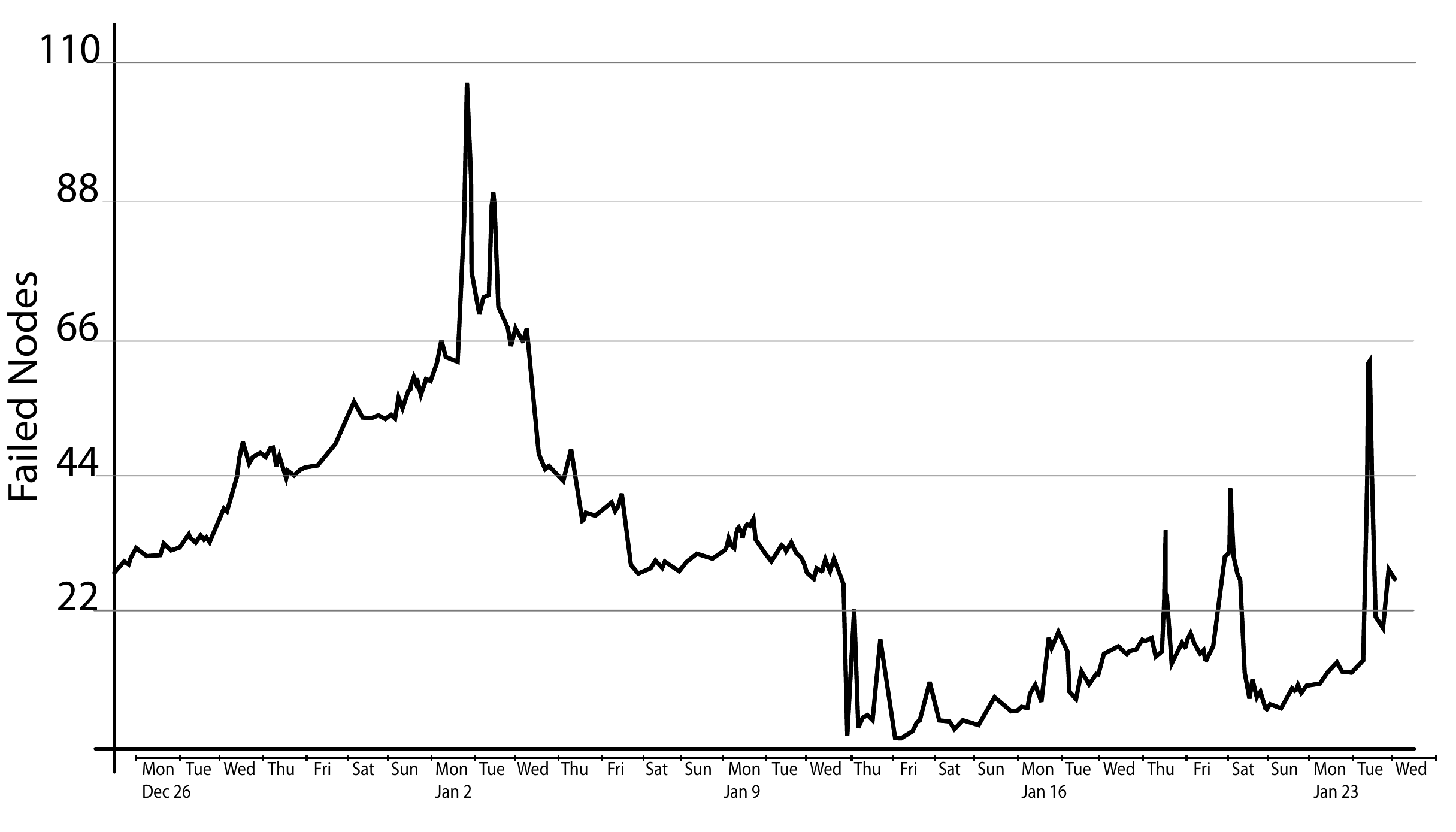}
\caption{Number of failed nodes over a single month period in a $3000$ node production cluster of Facebook.}
\label{fig:deadnodes}
\end{figure}

At Facebook, large analytics clusters store petabytes of information and handle multiple MapReduce analytics jobs. In a $3000$ node production cluster storing approximately $230$ million blocks (each of size $256$MB), only $8\%$ of the data is currently RS encoded (`RAIDed'). Fig.~\ref{fig:deadnodes} shows a recent trace of node failures in this production cluster. It is quite typical to have 20 or more node failures per day that trigger repair jobs, even when most repairs are delayed to avoid transient failures. 
A typical data node will be storing approximately $15$ TB and the repair traffic with the current configuration is estimated around $10-20\%$ of the total average of $2$ PB/day cluster network traffic. As discussed, (10,4) RS encoded blocks require approximately $10\times$ more network repair overhead per bit compared to replicated blocks. We estimate that if $50\%$ of the cluster was RS encoded, the repair network traffic would completely saturate the cluster network links. \textit{Our goal is to design more efficient coding schemes that would allow a large fraction of the data to be coded without facing this repair bottleneck. This would save petabytes of storage overheads and significantly reduce cluster costs}.

There are four additional reasons why efficiently repairable codes are becoming increasingly important in coded storage systems. The first is the issue of \textit{degraded reads}. Transient errors with no permanent data loss correspond to $90\%$ of data center failure events~\cite{ford2010availability,Khan_Fast12}. During the period of a transient failure event, block reads of a coded stripe will be \textit{degraded} if the corresponding data blocks are unavailable. In this case, the missing data block can be reconstructed by a repair process, which is not aimed at fault tolerance but at higher data availability. The only difference with standard repair is that the reconstructed block does not have to be written in disk. For this reason, efficient and fast repair can significantly improve data availability. 

The second is the problem of efficient \textit{node decommissioning}. Hadoop offers the decommission feature to retire a faulty data node. Functional data has to be copied out of the node before decommission, a process that is complicated and time consuming. Fast repairs allow to treat node decommissioning as a scheduled repair and start a MapReduce job to recreate the blocks without creating very large network traffic. 

The third reason is that repair influences the \textit{performance} of other concurrent MapReduce jobs. Several researchers have observed that the main bottleneck in MapReduce is the network~\cite{Orchestra}. As mentioned, repair network traffic is currently consuming a non-negligible fraction of the cluster network bandwidth. This issue is becoming more 
significant as the storage used is increasing disproportionately fast compared to network bandwidth in data centers. This increasing storage density trend emphasizes the importance of local repairs when coding is used. 

Finally, local repair would be a key in facilitating \emph{geographically distributed} file systems across data centers. Geo-diversity has been identified as one of the key future directions for improving latency and reliability~\cite{Costofcloud}. Traditionally, sites used to distribute data across data centers via replication. This, however, dramatically increases the total storage cost. Reed-Solomon codes across geographic locations at this scale would be completely impractical due to the high bandwidth requirements across wide area networks. Our work makes local repairs possible at a marginally higher storage overhead cost.  

Replication is obviously the winner in optimizing the four issues discussed, but requires a very large storage overhead. On the opposing tradeoff point, MDS codes have minimal storage overhead for a given reliability requirement, but suffer in repair and hence in all these implied issues. 
One way to view the contribution of this paper is a new intermediate point on this tradeoff, that sacrifices some storage efficiency to gain in these other metrics. 

The remainder of this paper is organized as follows: We initially present our theoretical results, the construction of Locally Repairable Codes and the information theoretic optimality results. We defer the more technical proofs to the Appendix. Section~\ref{sec:system} presents the HDFS-Xorbas architecture and Section~\ref{sec:availability} discusses a Markov-based reliability analysis. Section~\ref{sec:experimental} discusses our experimental evaluation on Amazon EC2 and Facebook's cluster. We finally 
survey related work in Section~\ref{sec:relatedwork} and conclude in Section~\ref{sec:conclusion}.

\section{Theoretical Contributions}

Maximum distance separable (MDS) codes are often used in various applications in communications and storage systems~\cite{Wicker}. A $(k,n-k)$-MDS code\footnote{In classical coding theory literature, codes are denoted by $(n,k)$ where $n$ is the number of data plus parity blocks, classically called blocklength. A (10,4) Reed-Solomon code would be classically denoted by RS (n=14,k=10). RS codes form the most well-known family of MDS codes. } of rate $R=\frac{k}{n}$ takes a file of size $M$, splits it in $k$ equally sized blocks, and then encodes it in $n$ coded blocks each of size $\frac{M}{k}$. Here we assume that our file has size exactly equal to $k$ data blocks to simplify the presentation; larger files are separated into stripes of $k$ data blocks and each stripe is coded separately. 

A $(k,n-k)$-MDS code has the property that any $k$ out of the $n$ coded blocks can be used to reconstruct the entire file.
It is easy to prove that this is the best fault tolerance possible for this level of redundancy: any set of $k$ blocks has an aggregate size of $M$ and therefore no smaller set of blocks could possibly recover the file.

Fault tolerance is captured by the metric of {\it minimum distance}.
\begin{defin}[Minimum Code  Distance] 
The minimum distance $d$ of a code of length $n$, is equal to the minimum number of erasures of coded blocks after which the file cannot be retrieved.
\end{defin}
MDS codes, as their name suggests, have the largest possible distance which is $d_{\text{MDS}}=n-k+1$. For example the minimum distance of a (10,4) RS is 
$n-k+1=5$ which means that five or more block erasures are needed to yield a data loss.

The second metric we will be interested in is  
{\it Block Locality}.
\begin{defin}[Block Locality]
An $(k,n-k)$ code has block locality $r$, when each coded block is a function of at most $r$ other coded blocks of the code.
\end{defin}

Codes with block locality $r$ have the property that, upon {\it any} single block erasure, fast repair of the lost coded block can be performed by computing a function on $r$ existing blocks of the code. This concept was recently and independently introduced in~\cite{Yekhanin,Oggier,papailiopoulos2011simple}.

When we require small locality, each single coded block should be repairable by using only a {\it small} subset of existing coded blocks $r<<k$, even when 
$n,k$ grow.
The following fact shows that locality and good distance are \textit{in conflict:} 
\begin{lem}
MDS codes with parameters $(k,n-k)$	cannot have locality smaller than $k$.
\label{MDS}
\end{lem}
Lemma \ref{MDS} implies that MDS codes have the {\it worst possible} locality since any $k$ blocks suffice to reconstruct the {\it entire} file, not just a single block. This is exactly the cost of optimal fault tolerance. 

The natural question is what is the best locality possible if we settled for ``almost MDS'' code distance. We answer this question and construct the first family of near-MDS codes with non-trivial locality. 
We provide a randomized and {\it explicit} family of codes that have {\it logarithmic} locality on all coded blocks and distance that is asymptotically equal to that of an MDS code. 
We call such codes $(k,n-k,r)$ Locally Repairable Codes (LRCs) and present their construction in the following section.
\begin{theo}
There exist $(k,n-k,r)$ Locally Repairable codes with logarithmic block locality $r=\log(k)$ and distance
$d_{\text{LRC}} = n-\left(1+\delta_k\right)k+1$.
Hence, any subset of $k\left(1+\delta_k\right)$ coded blocks can be used to reconstruct the file, where
$\delta_k = \frac{1}{\log(k)}-\frac{1}{k}$.
\label{theo:LRC}
\end{theo}

Observe that if we fix the code rate $R=\frac{k}{n}$ of an LRC and let $k$ grow, then its distance $d_{\text{LRC}}$ is almost that of a $(k,n-k)$-MDS code; hence the following corollary.
\begin{cor}
For fixed code rate $R = \frac{k}{n}$, the distance of LRCs is asymptotically equal to that of $(k,n-k)$-MDS codes
$$\lim_{k\rightarrow\infty}\frac{d_{\text{LRC}}}{d_{\text{MDS}}} = 1.$$
\end{cor}
LRCs are constructed on top of MDS codes (and the most common choice will be a Reed-Solomon code).

The MDS encoded blocks are grouped in logarithmic sized sets and then are combined together to obtain parity blocks of logarithmic degree. 
We prove that LRCs have the optimal distance for that specific locality, due to an information theoretic tradeoff that we establish.
Our locality-distance tradeoff is universal in the sense that it covers linear or nonlinear codes and is a generalization of recent result of Gopalan \textit{et al.}~\cite{Yekhanin} which established a similar bound for linear codes.
Our proof technique is based on building an information flow graph gadget, similar to the work of Dimakis~\textit{et al.}\cite{dimakis2010network,dimakis2011survey}. Our analysis can be found in the Appendix. 

\subsection{LRC implemented in Xorbas}

\begin{figure}[ht]
\centerline{\includegraphics[width=0.52\textwidth]{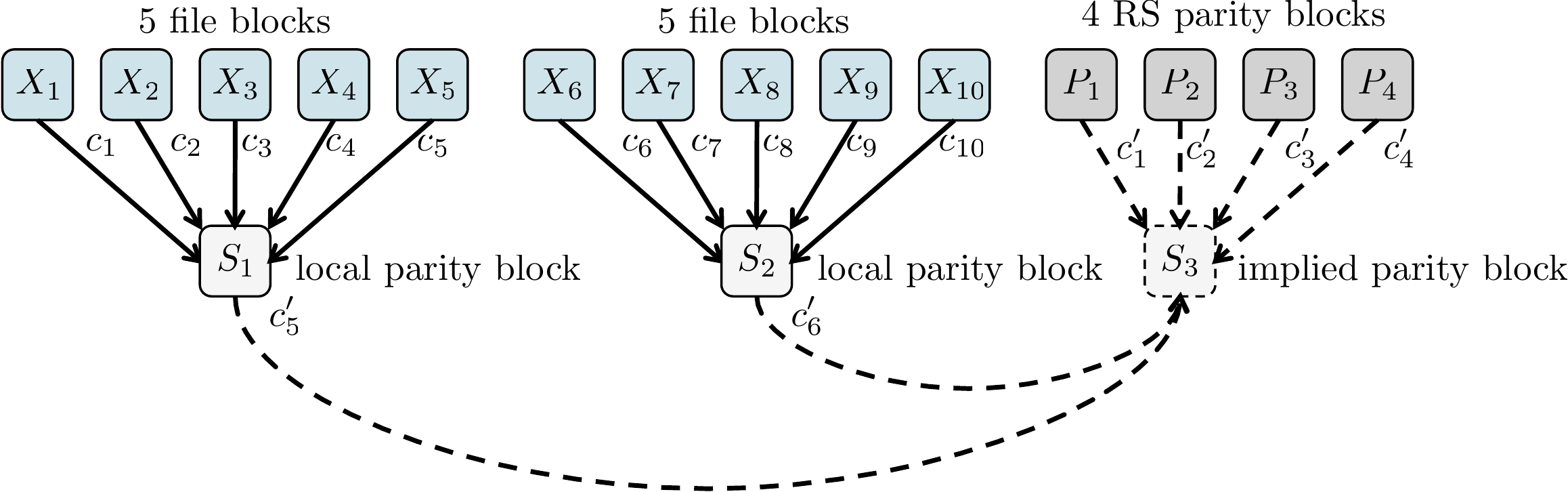}}
\caption{Locally repairable code implemented in HDFS-Xorbas. The four parity blocks $P_1, P_2, P_3, P_4$ are constructed with a standard RS code and the local parities provide efficient repair in the case of single block failures. 
The main theoretical challenge is to choose the coefficients $c_i$ to maximize the fault tolerance of the code. 
}
\label{fig:LRC16}
\end{figure}

We now describe the explicit $(10,6,5)$ LRC code we implemented in HDFS-Xorbas. For each stripe, we start with $10$ data blocks $X_1,X_2,\ldots ,X_{10}$ and use a $(10,4)$ Reed-Solomon 
over a binary extension field $\mathbb{F}_{2^m}$ to construct $4$ parity blocks $P_1,P_2,\ldots,P_4$. This is the code currently used in production clusters in Facebook that can tolerate any $4$ block failures due to the RS parities. The basic idea of LRCs is very simple: we make repair efficient by adding additional \emph{local parities}. 
This is shown in figure~\ref{fig:LRC16}. 

By adding the local parity $S_{1}= c_1 X_1+ c_2 X_2+c_3 X_3+c_4 X_5$, a single block failure can be repaired by accessing only $5$ other blocks. For example, if block $X_3$ is lost (or degraded read while unavailable) it can be reconstructed by
\begin{equation} 
	X_3= c_3^{-1} (S_{1}-c_1 X_1- c_2 X_2-c_4 X_4-c_5 X_5 ).
\end{equation}
The multiplicative inverse of the field element $c_3$ exists as long as $c_3 \neq 0$ which is the requirement we will enforce for all the local parity coefficients. It turns out that the coefficients $c_i$ can be selected to guarantee that all the linear equations will be linearly independent. In the Appendix we present a randomized and a deterministic algorithm to construct such coefficients. We emphasize that the complexity of the deterministic algorithm is exponential in the code parameters $(n,k)$ and therefore useful only for small code constructions. 

The disadvantage of adding these local parities is the extra storage requirement. While the original RS code was storing $14$ blocks for every $10$, the three local parities increase the storage overhead to $17/10$. There is one additional optimization that we can perform: We show that the coefficients $c_1,c_2, \ldots c_{10}$ can be chosen so that the local parities satisfy an additional
\textit{alignment equation} $S1+S2+S3=0$. We can therefore not store the local parity $S_3$ and 
instead consider it an \textit{implied parity}. Note that to obtain this in the figure, we set $c'_5=c'_6=1$.

When a single block failure happens in a RS parity, the implied parity can be reconstructed and used to repair that failure. For example, if $P_2$ is lost, it can be recovered by reading $5$ blocks
 $P_1,P_3,P_4, S_1,S_2$ and solving the equation 
\begin{equation} 
P_2= (c'_2)^{-1} (-S_1 -S_2 - c'_1 P_1 -c'_3 P_3-c'_4 P_4 ).
\end{equation}

In our theoretical analysis we show how to find non-zero coefficients $c_i$ (that must depend on the parities $P_i$ but are not data dependent) for the alignment condition to hold.
We also show that for the Reed-Solomon code implemented in HDFS RAID, choosing 
$c_i=1 \forall i$ and therefore performing simple XOR operations is sufficient. 
We further prove that this code has the largest possible distance ($d=5$) for this given locality $r=5$ and blocklength $n=16$.

\section{System Description}
\label{sec:system}
HDFS-RAID is an open source module that implements RS encoding and decoding over Apache Hadoop~\cite{hdfsraidwiki}. It provides a Distributed Raid File system (DRFS) that runs above HDFS. 
Files stored in DRFS are divided into stripes, \textit{i.e.}, groups of several blocks. 
For each stripe, a number of parity blocks are calculated and stored as a separate \textit{parity} file corresponding to the original file. HDFS-RAID is implemented in Java (approximately 12,000 lines of code) and is currently used in production by several organizations, including Facebook.

The module consists of several components, among which RaidNode and BlockFixer are the most relevant here:
\begin{itemize}

\item The RaidNode is a daemon responsible for the creation and maintenance of parity files for all data files stored in the DRFS.
One node in the cluster is generally designated to run the RaidNode. The daemon periodically scans the HDFS file system and decides whether a file is to be RAIDed or not, based on its size and age. In large clusters, RAIDing is done in a distributed manner by assigning MapReduce jobs to nodes across the cluster. 
After encoding, the RaidNode lowers the replication level of RAIDed files to one.  
\item The BlockFixer is a separate process that runs at the RaidNode and periodically checks for lost or corrupted blocks among the RAIDed files. 
When blocks are tagged as lost or corrupted, the BlockFixer rebuilds them using the surviving blocks of the stripe, again, by dispatching repair MapReduce (MR) jobs.
Note that these are not typical MR jobs.
Implemented under the MR framework, repair-jobs exploit its parallelization and scheduling properties, and can run along regular jobs under a single control mechanism.
\end{itemize}
Both RaidNode and BlockFixer rely on an underlying component: ErasureCode.
ErasureCode implements the erasure encoding/decoding functionality. In Facebook's HDFS-RAID, an RS $(10, 4)$ erasure code is implemented through ErasureCode ($4$ parity blocks are created for every $10$ data blocks). 


\subsection{HDFS-Xorbas}

Our system, {\bf HDFS-Xorbas} (or simply Xorbas), is a modification of HDFS-RAID that incorporates Locally Repairable Codes (LRC). To distinguish it from the HDFS-RAID implementing RS codes, we refer to the latter as {\bf HDFS-RS}. In Xorbas, the ErasureCode class has been extended to implement LRC on top of traditional RS codes. The RaidNode and BlockFixer classes were also subject to modifications in order to take advantage of the new coding scheme.

HDFS-Xorbas is designed for deployment in a large-scale Hadoop data warehouse, such as Facebook's clusters.
For that reason, our system provides backwards compatibility: Xorbas understands both LRC and RS codes 
and can incrementally modify RS encoded files into LRCs by adding only local XOR parities. 
To provide this integration with HDFS-RS, the specific LRCs we use are designed as extension codes of the (10,4) Reed-Solomon codes used at Facebook. First, a file is coded using RS code and then a small number of additional local parity blocks are created to provide local repairs.

\subsubsection{Encoding}
Once the RaidNode detects a file which is suitable for RAIDing (according to parameters set in a configuration file) it launches the encoder for the file. The encoder initially divides the file into stripes of $10$ blocks and calculates $4$ RS parity blocks.
Depending on the size of the file, the last stripe may contain fewer than $10$ blocks.
Incomplete stripes are considered as ``zero-padded`` full-stripes as far as the parity calculation is concerned

HDFS-Xorbas computes two extra parities for a total of $16$ blocks per stripe ($10$ data blocks, $4$ RS parities and $2$ Local XOR parities), as shown in Fig. \ref{fig:LRC16}.
Similar to the calculation of the RS parities, Xorbas calculates all parity blocks in a distributed manner through MapReduce encoder jobs.
All blocks are spread across the cluster according to Hadoop's configured block placement policy.
The default policy randomly places blocks at DataNodes, avoiding collocating blocks of the same stripe.

\subsubsection{Decoding \& Repair}

RaidNode starts a decoding process when corrupt files are detected. 
Xorbas uses two decoders: the light-decoder aimed at single block failures per stripe,
and the heavy-decoder, employed when the light-decoder fails. 

When the BlockFixer detects a missing (or corrupted) block, it determines the $5$ blocks required for the reconstruction according to the structure of the LRC.
A special MapReduce is dispatched to attempt light-decoding:
a single map task opens parallel streams to the nodes containing the required blocks, downloads them, and performs a simple XOR. 
In the presence of multiple failures, the $5$ required blocks may not be available.
In that case the light-decoder fails and the heavy decoder is initiated. 
The heavy decoder operates in the same way as in Reed-Solomon: streams to all the blocks of the stripe are opened and decoding is equivalent to solving a system of linear equations.
The RS linear system has a Vandermonde structure~\cite{Wicker} which allows small CPU utilization. 
The recovered block is finally sent and stored to a Datanode according to the cluster's block placement policy. 

In the currently deployed HDFS-RS implementation, even when a single block is corrupt, the BlockFixer 
opens streams to all $13$ other blocks of the stripe (which could be reduced to $10$ with a more efficient implementation). The benefit of Xorbas should therefore be clear: for all the single block failures and also many double block failures (as long as the two missing blocks belong to different local XORs), the network and disk I/O overheads will be significantly smaller.

\section{Reliability Analysis}
\label{sec:availability}
In this section, we provide a reliability analysis by estimating the mean-time to data loss (MTTDL)
using a standard Markov model.
We use the above metric and model to compare RS codes and LRCs to replication.
There are two main factors that affect the MTTDL: {\it i)} the number of block failures that we can tolerate before losing data and {\it ii)} the speed of block repairs. 
It should be clear that the MTTDL increases as the resiliency to failures increases and the time of block repairs decreases. 
In the following, we explore the interplay of these factors and their effect on the MTTDL.

When comparing the various schemes, replication offers the fastest repair possible at the cost of low failure resiliency. 
On the other hand, RS codes and LRCs can tolerate more failures, while requiring comparatively higher repair times, with the LRC requiring less repair time than RS. 
In \cite{ford2010availability}, the authors report values from Google clusters 
(cells) and show that, for their parameters, a $(9, 4)$-RS code has approximately six orders of magnitude higher reliability than 3-way replication.
Similarly here, we see how coding outperforms replication in terms of the reliability metric of interest.

Along with \cite{ford2010availability}, there exists significant work towards analyzing the reliability of replication, RAID storage~\cite{xin2003reliability}, and erasure codes~\cite{greenan2009reliability}. The main body of the above literature considers standard Markov models to analytically derive the MTTDL for the various storage settings considered. 
Consistent with the literature, we employ a similar approach to evaluate the reliability in our comparisons.
The values obtained here may not be meaningful in isolation but are useful for comparing the various schemes (see also~\cite{greenan2010mean}). 

In our analysis, the total cluster data is denoted by $C$ and $S$ denotes the stripe size. We set the number of disk nodes to be $N = 3000$, while the total data stored is set to be $C = 30$PB.
The mean time to failure of a disk node is set at $4$ years ($=1/\lambda$), and the block size is $B = 256$MB (the default value at Facebook's warehouses). 
Based on Facebook's cluster measurements, we limit the cross-rack communication to $\gamma=1$Gbps for repairs.
This limit is imposed to model the real cross-rack communication bandwidth limitations of the Facebook cluster.
In our case, the cross-rack communication is generated due to the fact that all coded blocks of a stripe are placed in different racks to provide higher fault tolerance.
This means that when repairing a single block, all downloaded blocks that participate in its repair are communicated across different racks.

Under 3-way replication, each stripe consists of  three blocks corresponding 
to the three replicas, and thus the total number of stripes in the system is 
${C}/{nB}$ where $n=3$. 
When RS codes or LRC is employed, the stripe size varies according to the code parameters $k$ and $n-k$.
For comparison purposes, we consider equal data stripe size $k=10$.
Thus, the number of stripes is ${C}/{nB}$, where $n = 14$ for (10, 4) RS and $n = 16$
for $(10, 6, 5)$-LRC. 
For the above values, we compute the MTTDL of a single stripe ($\text{MTTDL}_\text{stripe}$).
Then, we normalize the previous with the total number of stripes to get the MTTDL of the system, which is calculated as
\begin{equation}
\text{MTTDL} = \frac{\text{MTTDL}_\text{stripe}}{{C}/{nB}}.
\label{mttdleqn}
\end{equation}

\begin{figure}  
\centering
\includegraphics[width=0.9\columnwidth]{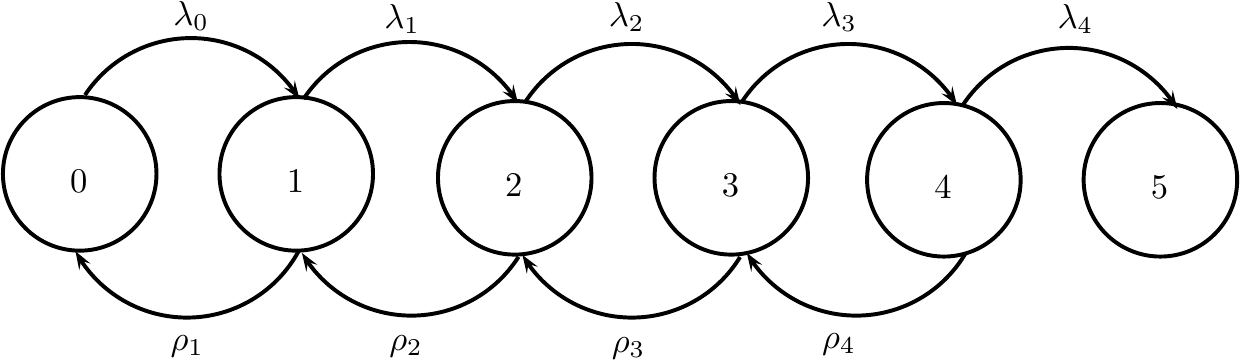}  
\caption{The Markov model used to calculate the $\text{MTTDL}_\text{stripe}$ of (10, 4) RS and
(10, 6, 5) LRC.}
\label{fig:LRCMarkov}
\end{figure}

Next, we explain how to compute the MTTDL of a stripe, for which we use a standard Markov model. 
The number of lost blocks at each time are used to denote the different states of the Markov chain. 
The failure and repair rates correspond to the forward and backward rates between the states.
When we employ 3-way replication, data loss occurs posterior to 3 block erasures.
For both the $(10, 4)$-RS and $(10, 6, 5)$-LRC schemes, 5 block erasures lead to data loss.
Hence, the Markov chains for the above storage scenarios will have a total of 3, 5, and 5 states, respectively.
In Fig.~\ref{fig:LRCMarkov}, we show the corresponding Markov chain for the $(10, 4)$-RS and the $(10, 6, 5)$-LRC. 
We note that although the chains have the same number of states, the transition probabilities will be different, depending on the coding scheme.

We continue by calculating the  transition rates. 
Inter-failure times are assumed to be exponentially distributed.
The same goes for the repair (backward) times.
In general, the repair times may not exhibit an exponential behavior, however, such an assumption simplifies our analysis.
When there are $i$ blocks remaining in a stripe (i.e., when the state is $n-i$), the rate at which a block is lost will 
be $\lambda_i = i\lambda$ because the $i$ blocks are distributed into different nodes and each
node fails independently at rate $\lambda$. 
The rate at which a block is repaired depends on how many blocks need to be 
downloaded for the repair, the block size, and the download rate $\gamma$.
For example, for the 3-replication scheme, single block repairs require downloading one block, hence we
assume $\rho_i = \gamma/B$, for $i = 1, 2$. 
For the coded schemes, we additionally consider the effect of using heavy or light decoders. 
For example in the LRC, if two blocks
are lost from the same stripe, we determine the probabilities for invoking light or heavy
decoder and thus compute the expected number of blocks to be downloaded. 
We skip a detailed derivation due to lack of space.
For a similar treatment, see~\cite{ford2010availability}.  
The stripe MTTDL equals the average time it takes to go from state $0$ to the ``data loss state". Under the above assumptions and transition rates, we calculate the MTTDL of the stripe from which the MTTDL of the system can be calculated using eqn~\ref{mttdleqn}.

The MTTDL values that we calculated  for replication, HDFS-RS, and Xorbas, under the Markov model considered, are shown in Table~\ref{table:MTTDLresults}. 
We observe that the higher repair speed of LRC compensates for the additional storage in terms of reliability.
This serves Xorbas LRC (10,6,5) two more zeros of reliability compared to a (10,4) Reed-Solomon code. 
The reliability of the 3-replication is substantially lower than both coded schemes, similar to what has been observed in related studies~\cite{ford2010availability}.

\begin{table}
\centering
\begin{tabular}{|c|c|c|c|}
\hline
 & Storage & Repair & MTTDL \\
Scheme & overhead & traffic & (days)\\
\hline
$3$-replication & 2x & 1x & $2.3079E+10$ \\
RS $(10, 4)$ & 0.4x & 10x & $3.3118E+13$ \\
LRC $(10, 6, 5)$ & 0.6x & 5x & $1.2180E+15$ \\
\hline
\end{tabular}
\caption{Comparison summary of the three schemes.
MTTDL assumes independent node failures.}
\label{table:MTTDLresults}
\end{table}

Another interesting metric is data availability. Availability is the fraction of 
time that data is available for use. Note that in the case of
3-replication, if one block is lost, then one of the other copies of the block is immediately available. 
On the contrary, for either RS or LRC, a job requesting a lost block must wait for the completion of the repair job. Since LRCs complete these jobs faster, they will have higher availability due to these faster degraded reads. 
A detailed study of availability tradeoffs of coded storage systems remains an interesting future research direction.

\section{Evaluation}
\label{sec:experimental}
\begin{figure*}[t!]
\subfloat[HDFS Bytes Read per failure event.]{\label{fig:50datanodes-200files-HDFS-BYTES-READ-per-event}
\includegraphics[width=0.33\textwidth]{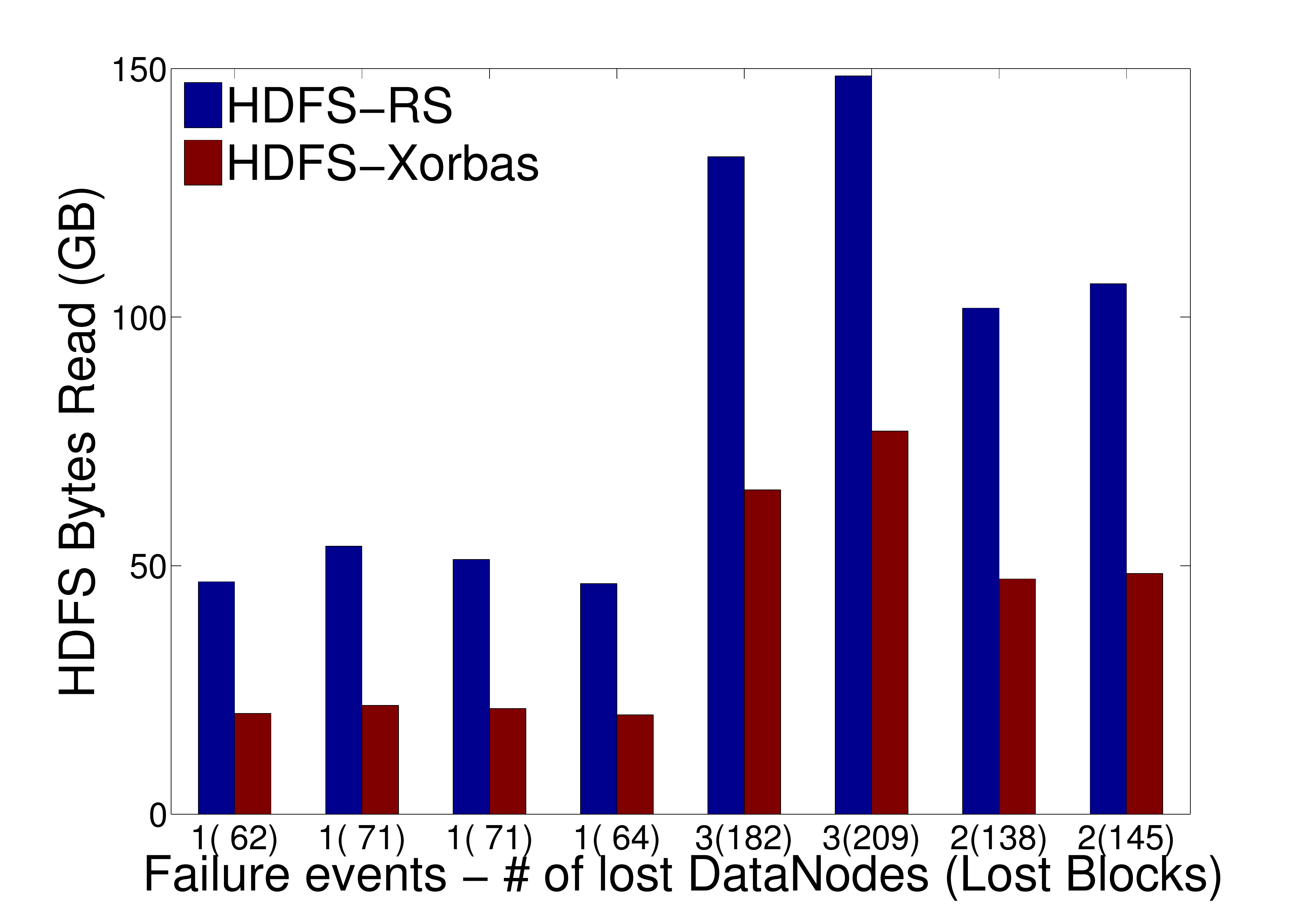}}
\subfloat[Network Out Traffic per failure event.]{\label{fig:50datanodes-200files-NETWORKOUT-per-event}
\includegraphics[width=0.33\textwidth]{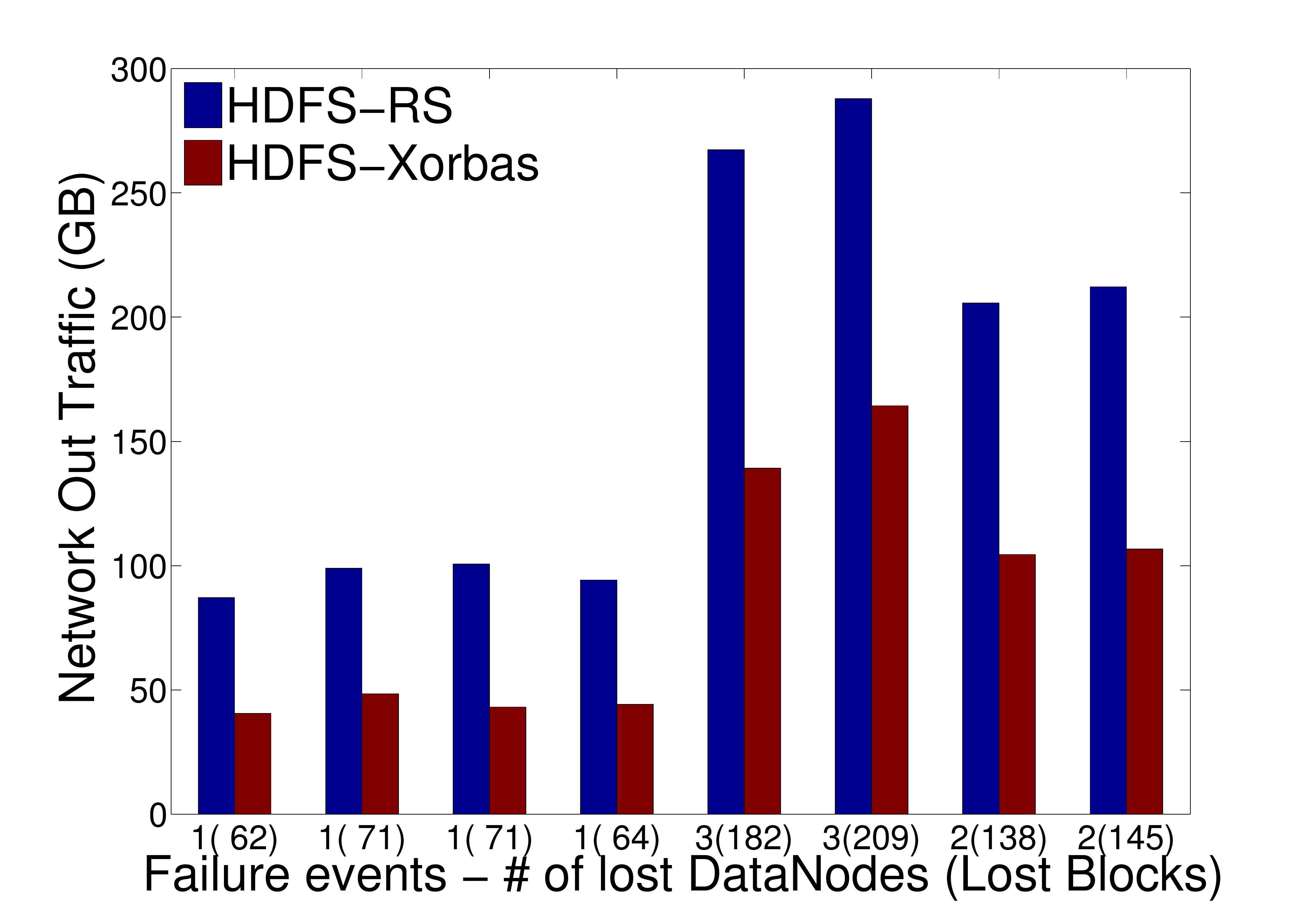}}
\subfloat[Repair duration per failure event.]{\label{fig:50datanodes-200files-REPAIR-DURATION-per-event}
\includegraphics[width=0.305\textwidth]{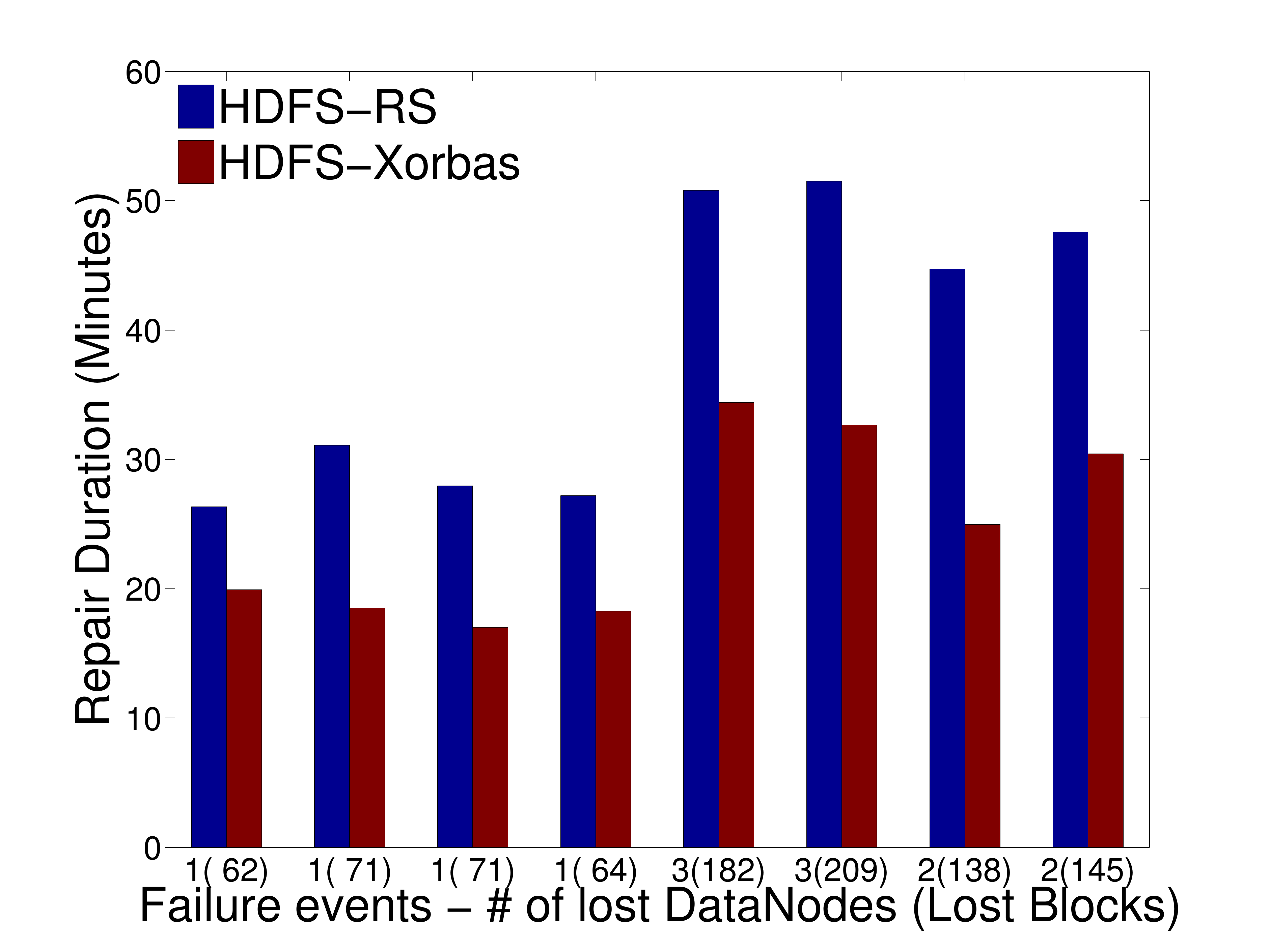}}
\caption{The metrics measured during the $200$ file experiment. Network-in is similar to Network-out and so it is not displayed here. During the course of the experiment, we simulated eight failure events and the x-axis gives details of the number of DataNodes terminated during each failure event and the number of blocks lost are displayed in parentheses. }
\label{fig:50datanodes-200files-measurements-per-event}
\end{figure*}
In this section, we provide details on a series of experiments we performed to evaluate the performance of HDFS-Xorbas in two environments: Amazon's Elastic Compute Cloud (EC2)~\cite{amazonec2} and a test cluster in Facebook.
\subsection{Evaluation Metrics}
We rely primarily on the following metrics to evaluate HDFS-Xorbas against HDFS-RS: HDFS Bytes Read, Network Traffic, and Repair Duration. 
HDFS Bytes Read corresponds to the total amount of data read by the jobs initiated for repair.
It is obtained by aggregating partial measurements collected from the statistics-reports of the jobs spawned following a failure event.
Network Traffic represents the total amount of data communicated from nodes in the cluster (measured in GB). 
Since the cluster does not handle any external traffic, Network Traffic is equal to the amount of data moving into nodes.
It is measured using Amazon's AWS Cloudwatch monitoring tools. 
Repair Duration is simply calculated as the time interval between the starting time of the first repair job and the ending time of the last repair job. 

\subsection{Amazon EC2}
On EC2, we created two Hadoop clusters, one running HDFS-RS and the other HDFS-Xorbas.
Each cluster consisted of $51$ instances of type m1.small, which corresponds to a $32$-bit machine with $1.7$ GB memory, $1$ compute unit and $160$ GB of storage, running Ubuntu/Linux-2.6.32. 
One instance in each cluster served as a master, hosting Hadoop's NameNode, JobTracker and RaidNode daemons, while the remaining $50$ instances served as slaves for HDFS and MapReduce, each hosting a DataNode and a TaskTracker daemon, thereby forming a Hadoop cluster of total capacity roughly equal to $7.4$ TB. 
Unfortunately, no information is provided by EC2 on the topology of the cluster. 

The clusters were initially loaded with the same amount of logical data.
Then a common pattern of failure events was triggered manually in both clusters to study the dynamics of data recovery.
The objective was to measure key properties such as the number of HDFS Bytes Read and the real Network Traffic generated by the repairs.

All files used were of size $640$ MB.
With block size configured to $64$ MB, each file yields a single stripe with $14$ and $16$ full size blocks in HDFS-RS and HDFS-Xorbas respectively.
We used a block size of $64$ MB, and all our files were of size $640$ MB. Therefore, each file yields a single stripe with $14$ and $16$ full size blocks in HDFS-RS and HDFS-Xorbas respectively. 
This choice is representative of the majority of stripes in a production Hadoop cluster: extremely large files are split into many stripes, so in total only a small fraction of the stripes will have a smaller size.
In addition, it allows us to better predict the total amount of data that needs to be read in order to reconstruct missing blocks and hence interpret our experimental results. 
Finally, since block repair depends only on blocks of the same stripe, using larger files that would yield more than one stripe would not affect our results. An experiment involving arbitrary file sizes, is discussed in Section \ref{sec:xcluster}.

During the course of a single experiment, once all files were RAIDed, a total of eight failure events were triggered in each cluster.
A failure event consists of the termination of one or more DataNodes. 
In our failure pattern, the first four failure events consisted of single DataNodes terminations, the next two were terminations of triplets of DataNodes and finally two terminations of pairs of DataNodes.
Upon a failure event, MapReduce repair jobs are spawned by the RaidNode to restore missing blocks.
Sufficient time was provided for both clusters to complete the repair process, allowing measurements corresponding to distinct events to be isolated.
For example, events are distinct in Fig. \ref{fig:50datanodes-200files-measurements-per-event}.
Note that the Datanodes selected for termination stored roughly the same number of blocks for both clusters.
The objective was to compare the two systems for the repair cost per block lost.
However, since Xorbas has an additional storage overhead, a random failure event would in expectation, lead to loss of $14.3$\% more blocks in Xorbas compared to RS. In any case, results can be adjusted to take this into account, without significantly affecting the gains observed in our experiments.

In total, three experiments were performed on the above setup, successively increasing the number of files stored ($50$, $100$, and $200$ files), in order to understand the impact of the amount of data stored on system performance. 
Fig.~\ref{fig:50datanodes-200files-measurements-per-event} depicts the measurement from the last case, while the other two produce similar results.
The measurements of all the experiments are combined in Fig. \ref{fig:linearleastsquares}, plotting HDFS Bytes Read, Network Traffic and Repair Duration versus the number of blocks lost, for all three experiments carried out in EC2. We also plot the linear least squares fitting curve for these measurements.

\subsubsection{HDFS Bytes Read}
Fig.~\ref{fig:50datanodes-200files-HDFS-BYTES-READ-per-event} depicts the total number of HDFS bytes read by the BlockFixer jobs initiated during each failure event. The bar plots show that HDFS-Xorbas reads $41\%-52\%$ the amount of data that RS reads to reconstruct the same number of lost blocks. These measurements are consistent with the theoretically expected values, given that more than one blocks per stripe are occasionally lost (note that $12.14/5= 41\%$). 
Fig.~\ref{fig:hdfsbytesread_linearleastsquares} shows that the number of HDFS bytes read is linearly dependent on the number of blocks lost, as expected. The slopes give us the average number of HDFS bytes read per block for Xorbas and HDFS-RS. The average number of blocks read per lost block 
are estimated to be $11.5$ and $5.8$, showing the $2\times$ benefit of HDFS-Xorbas. 

\subsubsection{Network Traffic}
\begin{figure*}[t!]
\centering
\subfloat[Cluster network traffic.]{\label{fig:50datanodes-200files-NetworkOut-in-time}\includegraphics[width=0.33\textwidth]{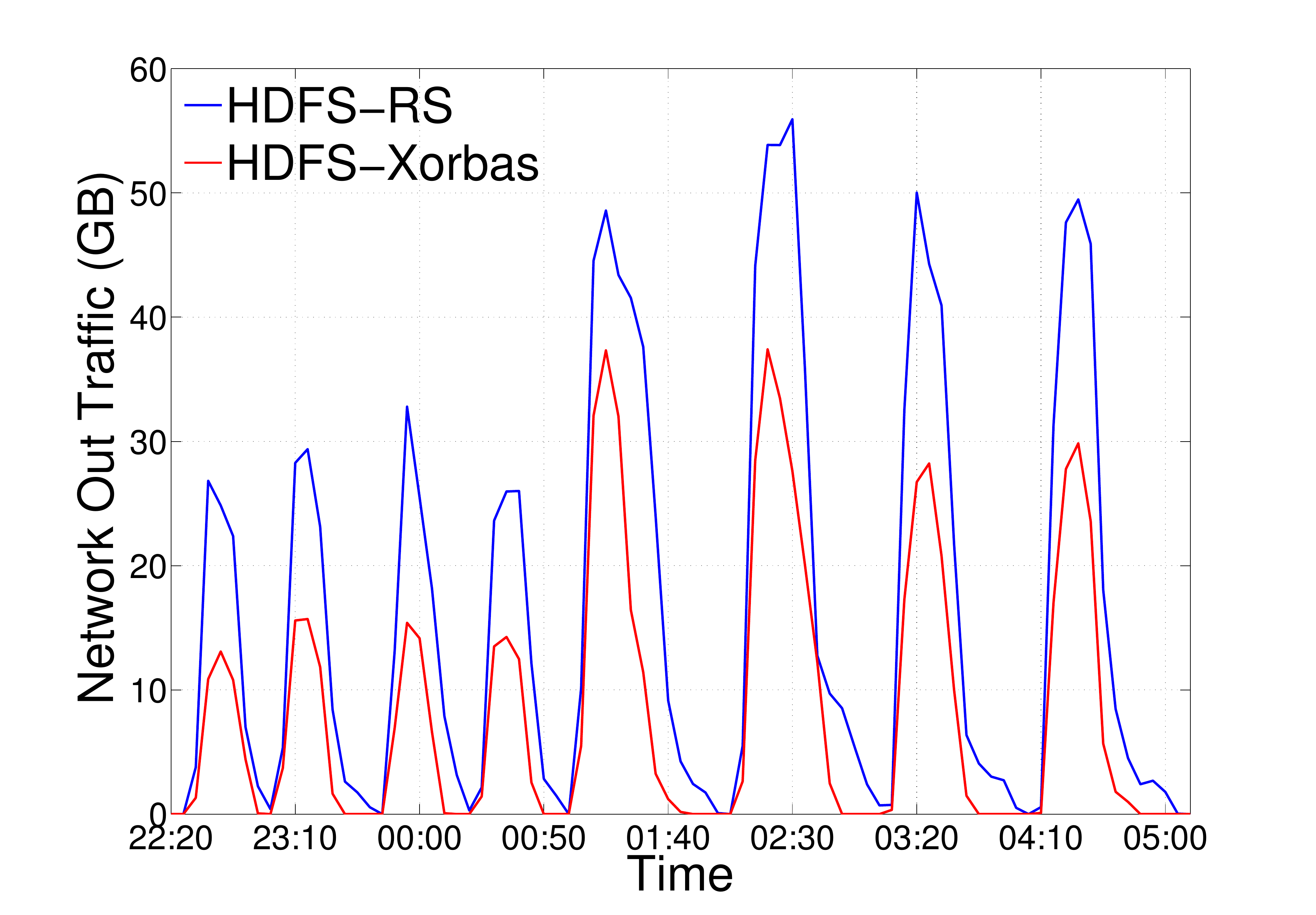}}
\subfloat[Cluster Disk Bytes Read.]{\label{fig:50datanodes-200files-DiskReadBytes-in-time}\includegraphics[width=0.33\textwidth]{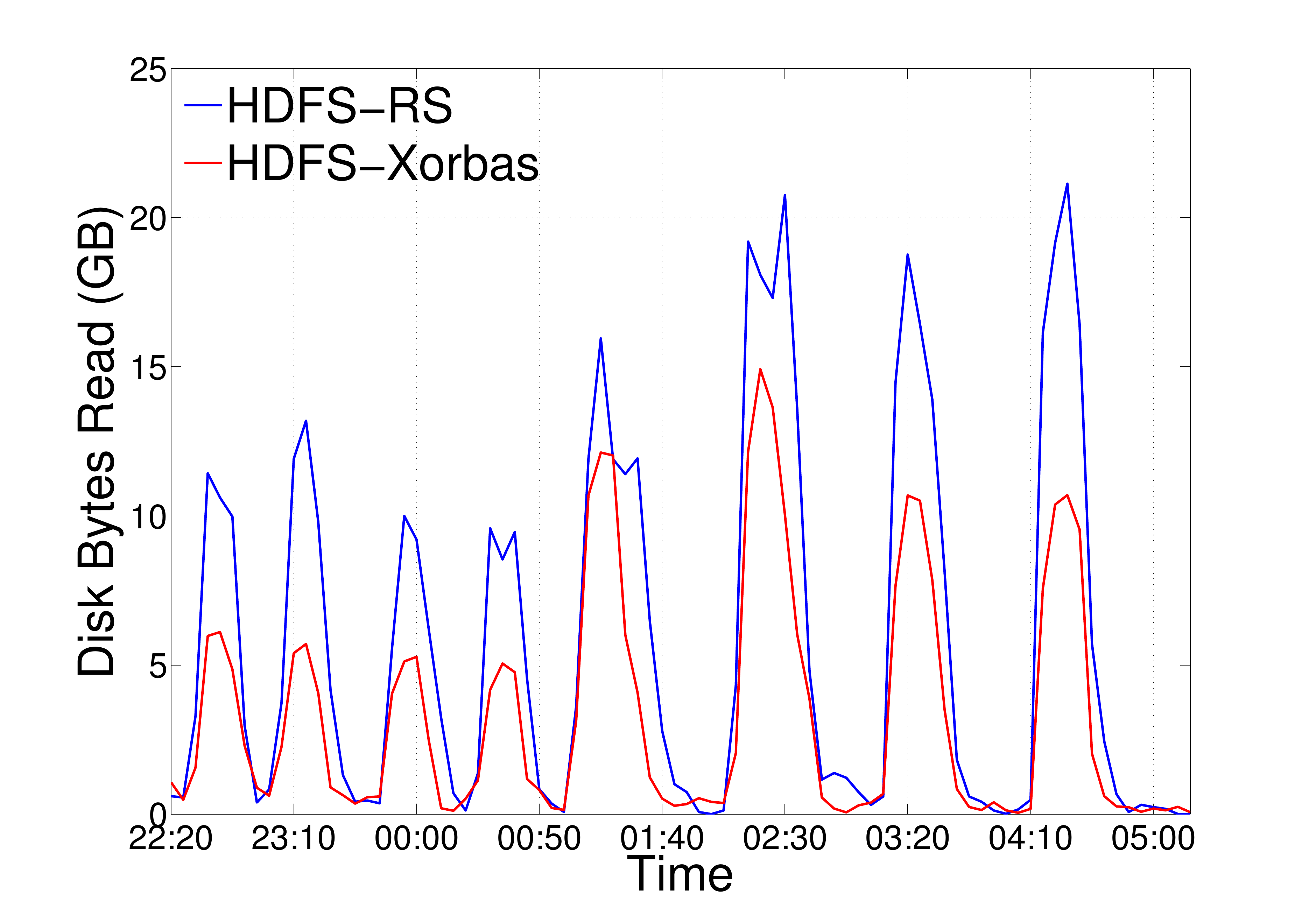}}
\subfloat[Cluster average CPU utilization.]{\label{fig:50datanodes-200files-CPUUtilization-in-time}\includegraphics[width=0.33\textwidth]{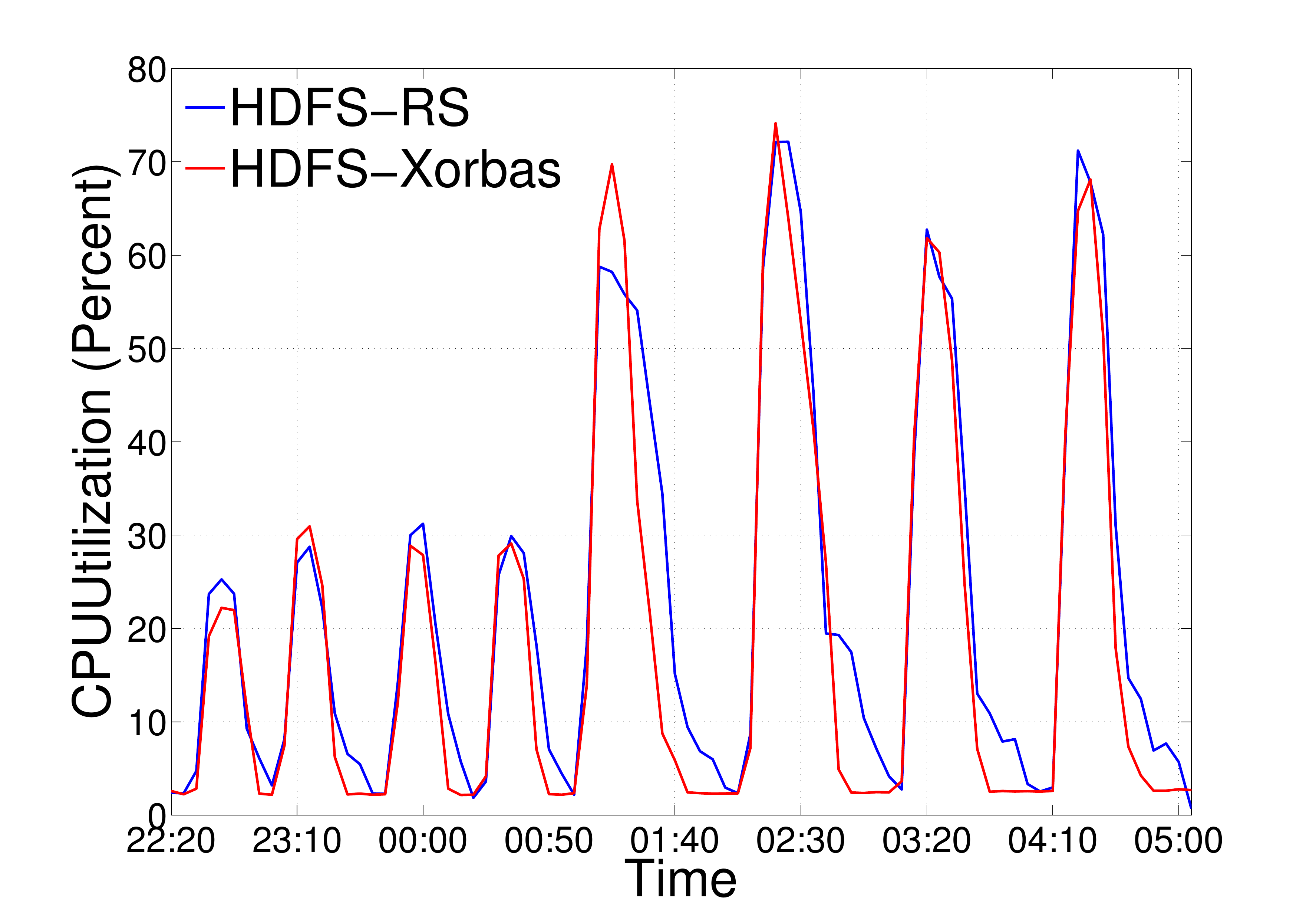}}
\caption{Measurements in time from the two EC2 clusters during the sequence of failing events. }
\label{fig:50datanodes-200files-measurements-in-time}
\end{figure*}

Fig.~\ref{fig:50datanodes-200files-NETWORKOUT-per-event} depicts the network traffic produced by BlockFixer jobs during the entire repair procedure. In particular, it shows the outgoing network traffic produced in the cluster, aggregated across instances. Incoming network traffic is similar since the cluster only communicates information internally.  In Fig.~\ref{fig:50datanodes-200files-NetworkOut-in-time}, we present the Network Traffic plotted continuously during the course of the 200 file experiment, with a 5-minute resolution. The sequence of failure events is clearly visible. Throughout our experiments, we consistently observed that network traffic was roughly equal to twice the number of bytes read. Therefore, gains in the number of HDFS bytes read  translate to network traffic gains, as expected. 

\subsubsection{Repair Time}
\begin{figure*}[t!]
\centering
 \subfloat[HDFS Bytes Read versus blocks lost]{\label{fig:hdfsbytesread_linearleastsquares}
 \includegraphics[width=0.33\textwidth]{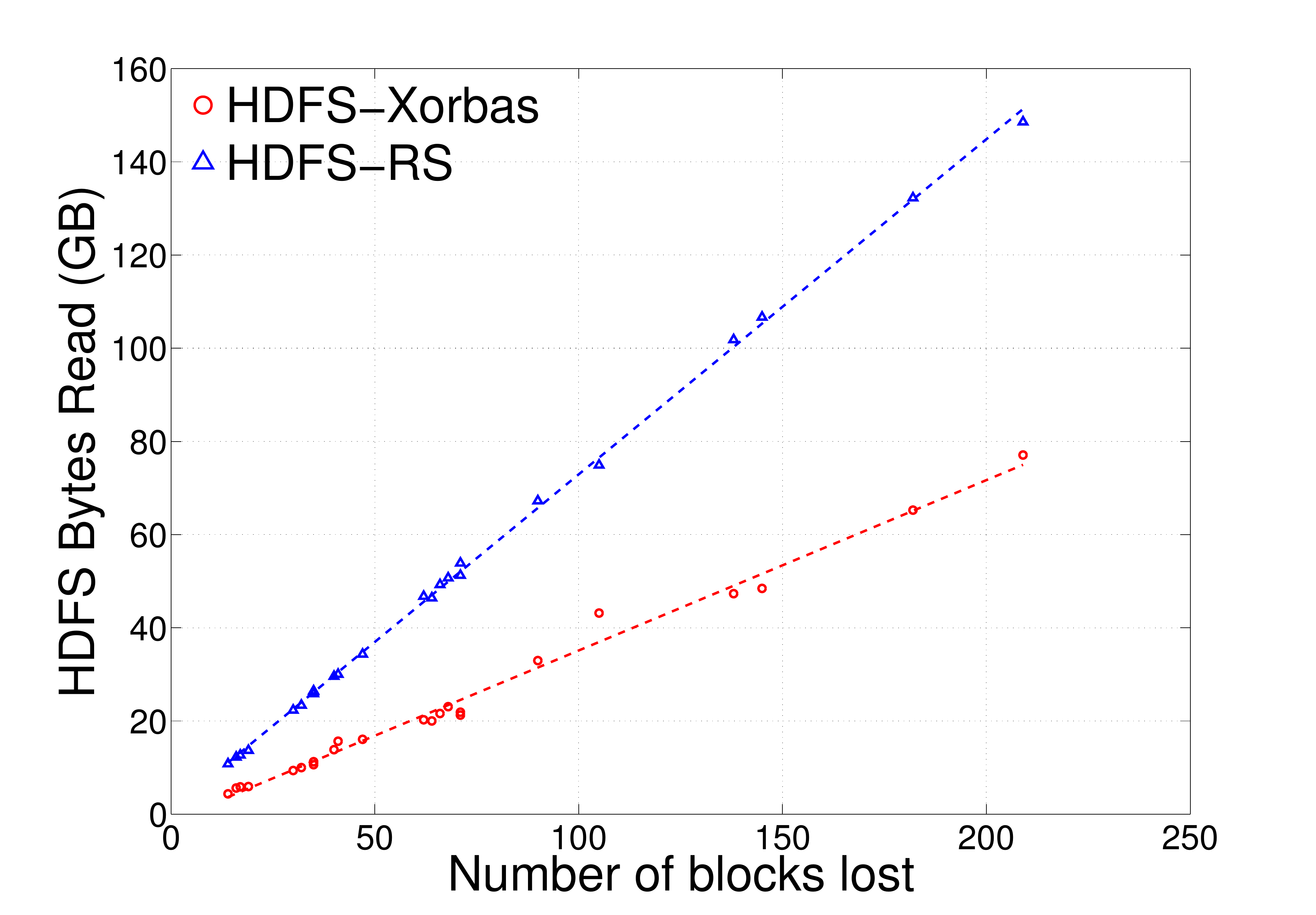}}
\subfloat[Network-Out Traffic]{\label{fig:networkout_linearleastsquares}
\includegraphics[width=0.33\textwidth]{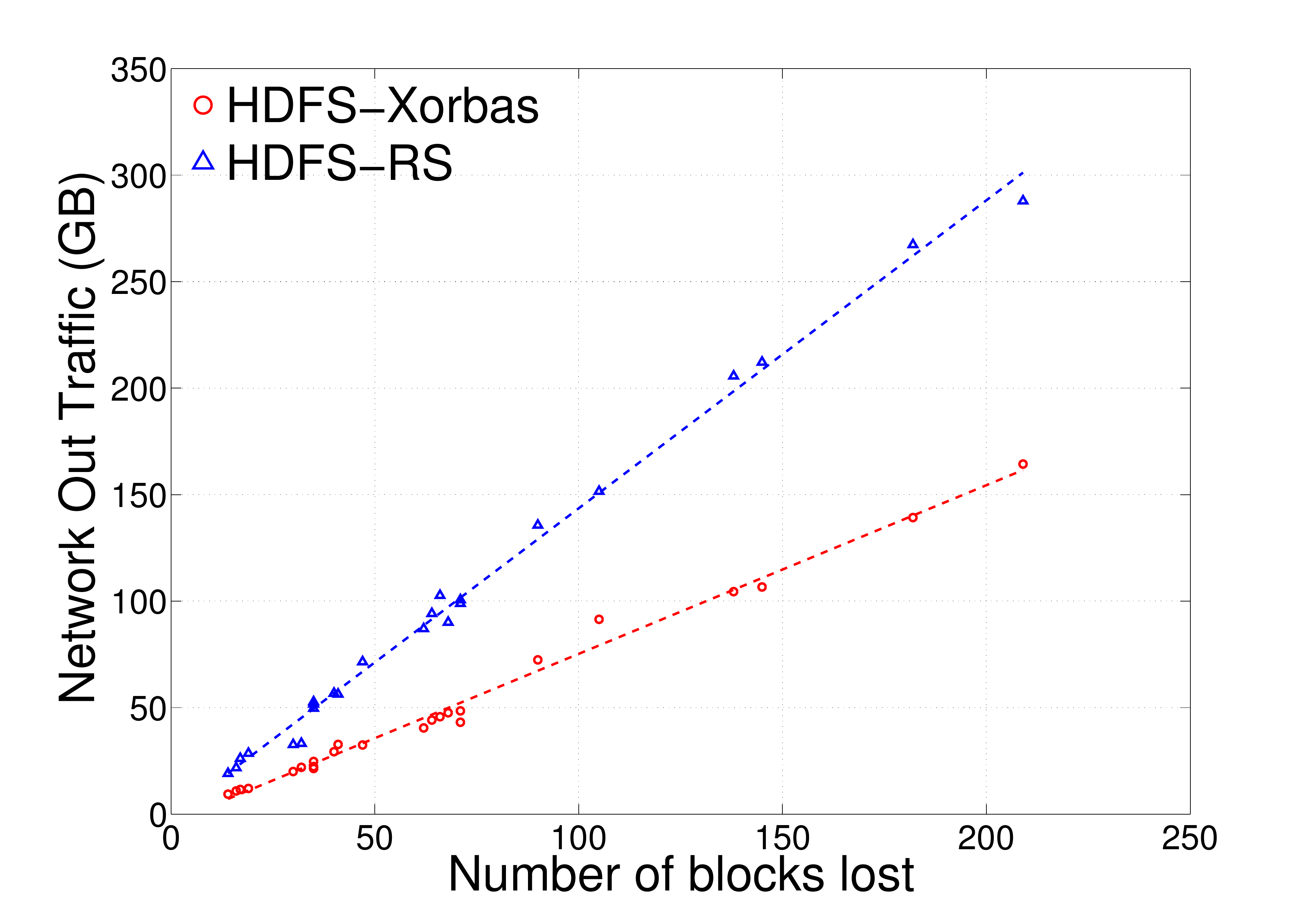}}
\subfloat[Repair Duration versus blocks lost]{\label{fig:repairtime_linearleastsquares} 
\includegraphics[width=0.33\textwidth]{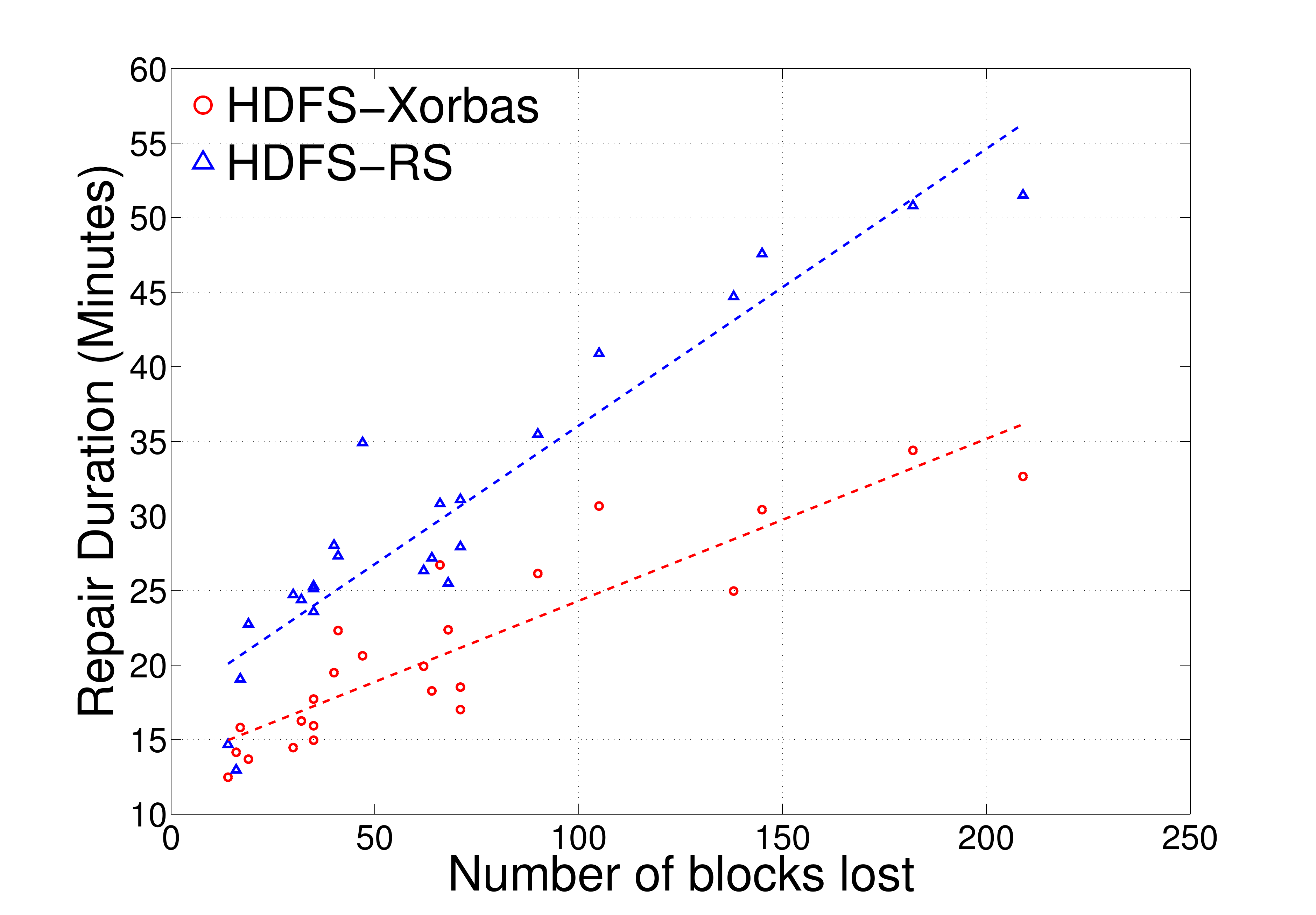}}
\caption{Measurement points of failure events versus the total number of blocks lost in the corresponding events. Measurements are from all three experiments. }
\label{fig:linearleastsquares}
\end{figure*}

Fig.~\ref{fig:50datanodes-200files-REPAIR-DURATION-per-event} depicts the total duration of the recovery procedure \textit{i.e.}, the interval from the launch time of the first block fixing job to the termination of the last one. Combining measurements from all the experiments, Fig.~\ref{fig:repairtime_linearleastsquares} shows the repair duration versus the number of blocks repaired. These figures show that Xorbas finishes $25\%$ to $45\%$ faster than HDFS-RS. 

The fact that the traffic peaks of the two systems are different is an indication that the available bandwidth was not fully saturated in these experiments. However, it is consistently reported that the network is typically the bottleneck for large-scale MapReduce tasks~\cite{Orchestra,VL2, DCell}. Similar behavior is observed in the Facebook production cluster at large-scale repairs. This is because hundreds of machines can share a single top-level switch which becomes saturated. Therefore, since LRC transfers significantly less data, we expect network saturation to further delay RS repairs in larger scale and hence give higher recovery time gains of LRC over RS.

From the CPU Utilization plots we conclude that HDFS RS and Xorbas have very similar CPU requirements and this
does not seem to influence the repair times. 

\subsubsection{Repair under Workload}
\begin{figure}
\centering
\includegraphics[width=0.5\textwidth]{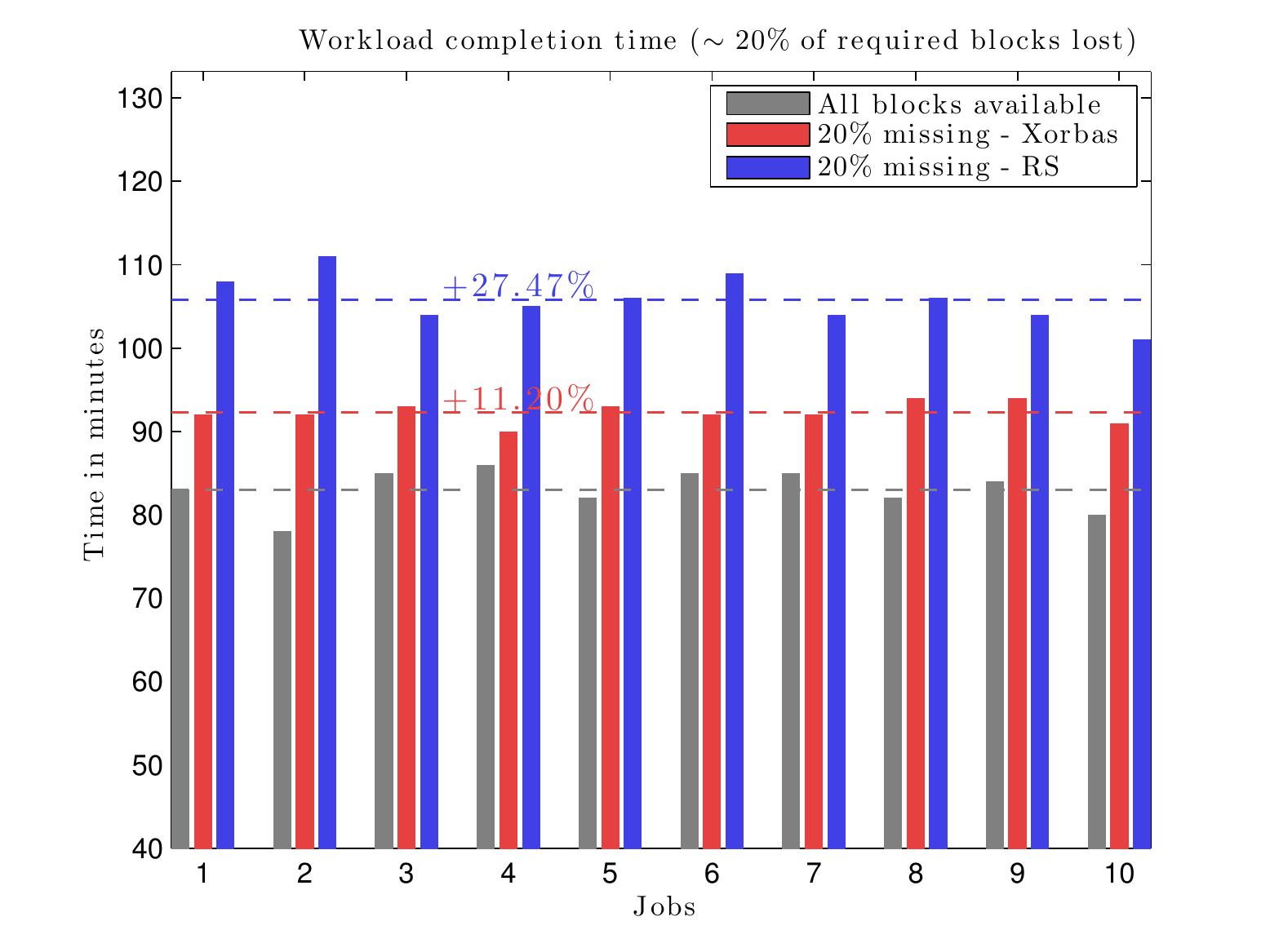}
\caption{Completion times of $10$ WordCount jobs: encountering no block missing, and $\sim 20\%$ of blocks missing on the two clusters. Dotted lines depict average job completion times.}
\label{fig:workload-jobs-completion-times}
\end{figure}

To demonstrate the impact of repair performance on the cluster's workload,
we simulate block losses in a cluster executing other tasks.
We created two clusters, $15$ slave nodes each.
The submitted artificial workload consists of word-count jobs running on five identical $3$GB text files.
Each job comprises several tasks enough to occupy all computational slots, while Hadoop's FairScheduler allocates tasks to TaskTrackers so that computational time is fairly shared among jobs.
Fig.~\ref{fig:workload-jobs-completion-times} depicts the execution time of each job under two scenarios: {\it i)} all blocks are available upon request, and {\it ii)} almost $20\%$ of the required blocks are missing.
Unavailable blocks must be reconstructed to be accessed, incurring a delay in the job completion which is much smaller in the case of HDFS-Xorbas. In the conducted experiments the additional delay due to missing blocks is more than doubled (from $9$ minutes for LRC to $23$ minutes for RS). 

We note that the benefits depend critically on how the Hadoop FairScheduler is configured. If concurrent jobs are blocked but the scheduler still allocates slots to them, delays can significantly increase. Further, jobs that need to read blocks may fail if repair times exceed a threshold. In these experiments we 
set the scheduling configuration options in the way most favorable to RS. 
Finally, as previously discussed, we expect that LRCs will be even faster than RS in larger-scale experiments due to network saturation. 
\begin{table}
\centering
\begin{tabular}[H]{c|c|c|c}
& All Blocks & \multicolumn{2}{c}{$\sim20\%$ of blocks missing}\\
& Avail.& RS & Xorbas \\ \hline
Total Bytes Read & $30$ GB& $43.88$ GB& $74.06$ GB\\
Avg Job Ex. Time & $83$ min& $92$ min& $106$ min\\
\end{tabular}
\caption{Repair impact on workload.}
\label{table:repair-under-workload}
\end{table}

\subsection{Facebook's cluster}
\label{sec:xcluster}
In addition to the series of controlled experiments performed over EC2, we performed one more experiment on Facebook's test cluster. This test cluster consisted of $35$ nodes configured with a total capacity of $370$ TB. 
Instead of placing files of pre-determined sizes as we did in EC2, we utilized the existing set of files in the cluster: $3,262$ files, totaling to approximately $2.7$ TB of logical data. The block size used was $256$ MB (same as in Facebook's production clusters). Roughly $94\%$ of the files consisted of $3$ blocks and the remaining of $10$ blocks, leading to an average $3.4$ blocks per file. 
\begin{table}[tbph!]
\centering
\begin{tabular}{l||c|c|c|c}
	 & Blocks&\multicolumn{2}{c|}{HDFS GB read}& Repair\\
	 & Lost    & Total & /block & Duration \\ \hline \hline
RS        &  $369$  &  $486.6$   &  $1.318$  & $26$ min\\
Xorbas &  $563$  &  $330.8$   &  $0.58$    & $19$ min
\end{tabular}
\caption{Experiment on Facebook's Cluster Results.}
\label{table:facebook-testcluster-experiment}
\end{table}

For our experiment, HDFS-RS was deployed on the cluster and upon completion of data RAIDing, a random DataNode was terminated. HDFS Bytes Read and the Repair Duration measurements were collected. Unfortunately, we did not have access to Network Traffic measurements. The experiment was repeated, deploying HDFS-Xorbas on the same set-up. Results are shown in Table~\ref{table:facebook-testcluster-experiment}. 
Note that in this experiment, HDFS-Xorbas stored $27\%$ more than HDFS-RS (ideally, the overhead should be $13\%$), due to the small size of the majority of the files stored in the cluster. As noted before, files typically stored in HDFS are large (and small files are typically archived into large HAR files). Further, it may be emphasized that the particular dataset used for this experiment is by no means representative of the dataset stored in 
Facebook's production clusters. 

In this experiment, the number of blocks lost in the second run, exceed those of the first run by more than the storage overhead introduced by HDFS-Xorbas. However, we still observe benefits in the amount of data read and repair duration, and the gains are even more clearer when normalizing by the number of blocks lost.

\section{Related Work}
\label{sec:relatedwork}

Optimizing code designs for efficient repair is a topic that has recently attracted significant attention due to its relevance to distributed systems. There is a substantial volume of work and we only try to give a high-level overview here. The interested reader can refer to~\cite{dimakis2011survey} and references therein. 
 
The first important distinction in the literature is between \textit{functional} and \textit{exact} repair. 
Functional repair means that when a block is lost, a different block is created that maintains the $(n,k)$ fault tolerance of the code. The main problem with functional repair is that when a systematic block is lost, it will be replaced with a parity block. While global fault tolerance to $n-k$ erasures remains, reading a single block 
would now require access to $k$ blocks. While this could be useful for archival systems with rare reads, it is not practical for our workloads. Therefore, we are interested only in codes with \textit{exact} repair so that we can maintain the code systematic. 

Dimakis \textit{et al.}~\cite{dimakis2010network} showed that it is possible to repair codes with network traffic smaller than the naive scheme that reads and transfers $k$ blocks. The first regenerating codes~\cite{dimakis2010network} provided only functional repair and the existence of exact regenerating codes matching the information theoretic bounds remained open. 

A substantial volume of work  (e.g. \cite{dimakis2011survey,Rashmi1,Tamo} and references therein) subsequently showed that exact repair is possible, matching the information theoretic bound of~\cite{dimakis2010network}. The code constructions are separated into exact codes for low rates $k/n \leq 1/2$ and high rates $k/n>1/2$. For rates below $1/2$ (\textit{i.e.} storage overheads above $2$) beautiful combinatorial constructions of exact regenerating codes were recently discovered~\cite{rashmi2011optimal, shah2012interference}. 
Since replication has a storage overhead of three, for our applications storage overheads around $1.4-1.8$ are of most interest, which ruled out the use of low rate exact regenerating codes. 

For high-rate exact repair, our understanding is currently incomplete.
The problem of existence of such codes remained open until two groups independently~\cite{cadambeAsymptotic}  used \textit{Interference Alignment}, an asymptotic technique developed for wireless information theory, to show the existence of exact regenerating codes at rates above $1/2$. 
Unfortunately this construction is only of theoretical interest since it requires exponential field size and performs well only in the asymptotic regime. Explicit high-rate regenerating codes are a topic of active research but no practical construction is currently known to us. 
A second related issue is that many of these codes reduce the repair network traffic but at a cost of higher disk I/O. It is not currently known if this high disk I/O is a fundamental requirement or if practical codes with both small disk I/O and repair traffic exist.  
 
Another family of codes optimized for repair has focused on relaxing the MDS requirement to improve on repair disk I/O and network bandwidth (e.g.~\cite{Pyramid,kbp:11:iso,Yekhanin}). The metric used in these constructions is 
\textit{locality}, the number of blocks that need to be read to reconstruct a lost block.  
The codes we introduce are optimal in terms of locality and match the bound shown in~\cite{Yekhanin}. 
In our recent prior work~\cite{DimitrisISIT} we generalized this bound and showed that it is information theoretic
(i.e. holds also for vector linear and non-linear codes). We note that optimal locality does not necessarily mean 
optimal disk I/O or optimal network repair traffic and the fundamental connections of these quantities remain open. 

The main theoretical innovation of this paper is a novel code construction with optimal locality that relies on Reed-Solomon global parities. We show how the concept of implied parities can save storage and show how to explicitly achieve parity alignment if the global parities are Reed-Solomon.


\section{Conclusions}
\label{sec:conclusion}
Modern storage systems are transitioning to erasure coding. We introduced a new family of codes called Locally Repairable Codes (LRCs) that have marginally suboptimal storage but significantly smaller repair disk I/O and network bandwidth requirements. In our implementation, we observed $2\times$ disk I/O and network reduction for the cost of $14\%$ more storage, a price that seems reasonable for many scenarios.

One related area where we believe locally repairable codes can have a significant impact is purely archival clusters. In this case we can deploy large LRCs (\textit{i.e.}, stipe sizes of 50 or 100 blocks) that can simultaneously offer high fault tolerance and small storage overhead.  This would be impractical if Reed-Solomon codes are used since the repair traffic grows linearly in the stripe size. 
Local repairs would further allow spinning disks down~\cite{Writeoff} since very few are required for single block repairs. 

In conclusion, we believe that LRCs create a new operating point that will be practically relevant in large-scale storage systems, especially when the network bandwidth is the main performance bottleneck.

\begin{appendix}

\section{Distance and Locality through Entropy}

In the following, we use a characterization of the code distance  $d$ of a length $n$ code that is based on the entropy function. 
This characterization is universal in the sense that it covers any linear or nonlinear code designs.

Let ${\bf x}$ be a file of size $M$ that we wish to split and store with redundancy $\frac{k}{n}$ in $n$ blocks, where each block has size $\frac{M}{k}$.
Without loss of generality, we assume that the file is split in $k$ blocks of the same size
${\bf x}\eqdef [X_1\ldots X_k]\in\mathbb{F}^{1\times k}$,
where $\mathbb{F}$ is the finite field over which all operations are performed.
The entropy of each file block is
$H(X_i) = \frac{M}{k}$,
 for all $i\in[k]$, where $[n]=\{1,\ldots,n\}$.\footnote{Equivalently, each block can be considered as a random variable that has entropy $\frac{M}{k}$.}
Then, we define an encoding (generator) map $G:\mathbb{F}^{1\times k}\mapsto\mathbb{F}^{1\times n}$ that takes as input the $k$ file blocks and outputs $n$ coded blocks
$G({\bf x}) = {\bf y} = [Y_1\ldots Y_n]$,
where $H(Y_i)=\frac{M}{k}$, for all $i\in[n]$.
The encoding function $G$ defines a $(k,n-k)$ code $\mathcal{C}$ over the vector space $\mathbb{F}^{1\times n}$.
We can calculate the effective rate of the code as the ratio of the entropy of the file blocks to the sum of the entropies of the $n$ coded blocks
\begin{equation}
R = \frac{H(X_1,\ldots,X_k)}{\sum_{i=1}^n H(Y_i)}= \frac{k}{n}.
\end{equation}

The distance $d$ of the code $\mathcal{C}$ is equal to the minimum number of erasures of blocks in ${\bf y}$ after which the entropy of the remaining blocks is strictly less than $M$
\begin{equation}
d = \min_{H(\{Y_1,\ldots,Y_n\}\backslash \mathcal{E})<M}|\mathcal{E}|=n-\max_{H(\mathcal{S})<M}|\mathcal{S}|,
\end{equation}
where  $\mathcal{E} \in 2^{\{Y_1,\ldots, Y_n\}}$ is a block erasure pattern set and $2^{\{Y_1,\ldots, Y_n\}}$ denotes the power set of $\{Y_1,\ldots, Y_n\}$, i.e., the set that consists of all subset of $\{Y_1,\ldots, Y_n\}$.
Hence, for a code $\mathcal{C}$ of length $n$ and distance $d$, any $n-d+1$ coded blocks can reconstruct the file, i.e., have joint entropy at least equal to $M$.
It follows that when $d$ is given, $n-d$ is the maximum number of coded variables that have entropy less than $M$.

The locality $r$ of a code can also be defined in terms of coded block entropies.
When a coded block $Y_i$,  $i\in[n]$, has locality $r$, then it is a function of $r$ other coded variables 
$Y_i = f_i (Y_{\mathcal{R}(i)})$,
where $\mathcal{R}(i)$ indexes the set of $r$ blocks $Y_j$, $j\in\mathcal{R}(i)$, that can reconstruct $Y_i$, and $f_i$ is some function (linear or nonlinear) on these $r$ coded blocks.
Hence, the entropy of $Y_i$ conditioned on its repair group $\mathcal{R}(i)$ is identically equal to zero
$H(Y_i |f_i (Y_{\mathcal{R}(i)}))=0$,
for $i\in[n]$.
This functional dependency of $Y_i$ on the blocks in $\mathcal{R}(i)$ is fundamentally the only code structure that we assume in our derivations.\footnote{In the following, we consider codes with uniform locality, i.e., $(k,n-k)$ codes  where  all encoded blocks have locality $r$. These codes are referred to as non-canonical codes in~\cite{Yekhanin}.}
This generality is key to providing universal information theoretic bounds on the code distance of $(k,n-k)$ linear, or nonlinear, codes that have locality $r$. 
Our following bounds can be considered as generalizations of the Singleton Bound on the code distance when locality is taken into account.

\section{Information theoretic limits of Locality and Distance}
We consider $(k,n-k)$ codes that have block locality $r$.
We find a lower bound on the distance by lower bounding the largest set $\mathcal{S}$ of coded blocks whose entropy is less than $M$, i.e., a set that cannot reconstruct the file. 
Effectively, we solve the following optimization problem that needs to be performed over all possible codes $\mathcal{C}$ and yields a best-case minimum distance
\begin{align}
\min_{\mathcal{C}} \max_{\mathcal{S}}|\mathcal{S}|\text{ s.t.: }&H(\mathcal{S})<M,\;\;\mathcal{S}\in 2^{\{Y_1,\ldots, Y_n\}}.\nonumber
\end{align}
We are able to provide a bound by considering a single property: each block is a member of a repair group of size $r+1$.
\begin{defin}
For a code $\mathcal{C}$ of length $n$ and locality $r$, a coded block $Y_i$ along with the blocks that can generate it, $Y_{\mathcal{R}(i)}$, form a repair group
$\Gamma(i) = \left\{i,\mathcal{R}(i)\right\}$,
for all $i\in[n]$. We refer to these repair groups, as $(r+1)$-groups.
\end{defin}
It is easy to check that the joint entropy of the blocks in a single $(r+1)$-group is at most as much as the entropy of $r$ file blocks
\begin{align}
H\left(Y_{\Gamma(i)}\right)&=H\left(Y_i,Y_{\mathcal{R}(i)}\right)=H\left(Y_{\mathcal{R}(i)}\right)+H\left(Y_i|Y_{\mathcal{R}(i)}\right)\nonumber\\
&=H\left(Y_{\mathcal{R}(i)}\right)\le \sum_{j\in\mathcal{R}(i)}H(Y_j)= r\frac{M}{k},\nonumber
\end{align}
for all $i\in[n]$.
To determine the upper bound on minimum distance of $\mathcal{C}$, we
construct the maximum set of coded blocks $\mathcal{S}$ that has entropy less than $M$.
We use this set to derive the following theorem.
\begin{theo}
\label{theo:distance}
For a code $\mathcal{C}$ of length $n$, where each coded block has entropy $\frac{M}{k}$ and locality $r$, the minimum distance is bounded as
\begin{equation}
d\le n-\left\lceil\frac{k}{r}\right\rceil-k+2.
\end{equation}
\end{theo}
{\bf Proof}: Our proof follows the same steps as the one in \cite{Yekhanin}.
We start by building the set $\mathcal{S}$ in steps and denote the collection of coded blocks at each step as $\mathcal{S}_i$.
The algorithm that builds the set is in Fig. \ref{fig:buildS}.
The goal is to lower bound the cardinality of $\mathcal{S}$, which results in an upper bound on code distance $d$, since $d\le n-|\mathcal{S}|$.
At each step we denote the difference in cardinality of $\mathcal{S}_i$ ans $\mathcal{S}_{i-1}$ and the difference in entropy as
$s_i = |\mathcal{S}_i|-|\mathcal{S}_{i-1}| \text{ and }h_i = H(\mathcal{S}_i)-H(\mathcal{S}_{i-1})$,
respectively.

\begin{figure}[h]
\begin{center}
\begin{tabular}{|c|l|}
\hline
step &\\
\hline
1 & Set $\mathcal{S}_0=\emptyset$ and $i=1$\\
2 & \texttt{WHILE} $H(\mathcal{S}_{i-1})<M$\\
3 & \hspace{0.5cm}Pick a coded block $Y_j\notin\mathcal{S}_{i-1}$\\
4 & \hspace{0.5cm}\texttt{IF} $H(\mathcal{S}_{i-1}\cup\{Y_{\Gamma(j)}\})<M$\\
5& \hspace{1cm}set $\mathcal{S}_i=\mathcal{S}_{i-1}\cup Y_{\Gamma(j)}$\\
6& \hspace{0.5cm}\texttt{ELSE IF} $H(\mathcal{S}_{i-1}\cup\{Y_{\Gamma(j)}\})\ge M$ \\
7& \hspace{1cm}pick $\mathcal{Y}_s\subset  Y_{\Gamma(j)} $ s.t. $H(\mathcal{Y}_s\cup\mathcal{S}_{i-1})<M$\\
8 &\hspace{1cm}set $\mathcal{S}_i=\mathcal{S}_{i-1}\cup\mathcal{Y}_s$\\
9& \hspace{0.5cm}$i=i+1$\\
\hline
\end{tabular}
\end{center}
\caption{The algorithm that builds set $\mathcal{S}$.}
\label{fig:buildS}
\end{figure}

At each step (depending on the possibility that two $(r+1)$-groups overlap) the difference in cardinalities $s_i$ is bounded as
$1\le s_i\le r+1$,
that is $s_i=r+1-p$, where $\left|\{Y_{\Gamma(j)}\}\cap \mathcal{S}_{i-1}\right|=p$.
Now there exist two possible cases.
First, the case where the last step set $\mathcal{S}_l$ is generated by line 5.
For this case we can also bound the entropy as
$h_i\le (s_i-1)\frac{M}{k}\Leftrightarrow s_i\ge \frac{k}{M}h_i+1$
which comes from the fact that, at least one coded variable in $\{Y_{\Gamma(j)}\}$ is a function of variables in
$\mathcal{S}_{i-1}\cup Y_{\mathcal{R}(j)}$.
Now, we can bound the cardinality
$\left|\mathcal{S}_l\right| =\sum_{i=1}^ls_i \ge \sum_{i=1}^l\left(\frac{kh_i}{M}+1\right)=l+\frac{k}{M}\sum_{i=1}^l 
h_i$.
We now have to bound $l$ and $\sum_{i=1}^l h_i$.
First, observe that since $l$ is our ``last step,'' this means that the aggregate entropy in $\mathcal{S}_l$ should be less than the file size, i.e., it should have a value  $M-c\cdot \frac{M}{k}$, for $0<c\le 1$. If $c>1$ then we could collect another variable in that set.
On the other hand, if $c=0$, then the coded blocks in $\mathcal{S}_l$ would have been sufficient to reconstruct the file.
Hence, 
 $M-\frac{M}{k}\le \sum_{i=1}^l h_i <M$.
We shall now lower bound $l$.
The smallest $l'\le l$ (i.e., the fastest) upon which $\mathcal{S}_{l'}$ reaches an aggregate entropy that is greater than, or equal to $M$, can be found in the following way: if we could only collect $(r+1)$-groups of entropy $r\frac{M}{k}$, without ``entropy losses'' between these groups, i.e., if there were no further dependencies than the ones dictated by  locality, then we would stop just before  $\mathcal{S}_{l'}$ reached an entropy of $M$, that is
$\sum_{i=1}^{l'}h_{l'}<M \Leftrightarrow l'r\frac{M}{k} < M\Leftrightarrow l'<\left\lceil\frac{k}{r}\right\rceil$.
However, $l'$ is an integer, hence
$l'=\left\lceil\frac{k}{r}\right\rceil-1$.
We apply the above to bound the cardinality
$|\mathcal{S}_{l}|\ge k-1+l'\ge k-1+\left\lceil\frac{k}{r}\right\rceil-1=k+\left\lceil\frac{k}{r}\right\rceil-2$,
in which case we obtain
$d\le n-\left\lceil\frac{k}{r}\right\rceil-k+2$.

We move to the second case where we reach line 6 of the building algorithm: the entropy of the file can be covered only by collecting $(r+1)$ groups.
This depends on the remainder of the division of $M$ by $r\frac{M}{k}$.
Posterior to collecting the $(r+1)$-groups, we are left with some entropy that needs to be covered by at most $r-1$ additional blocks not in $\mathcal{S}_{l'}$.
The entropy not covered by the set $\mathcal{S}_{l'}$ is
$M-l'r\frac{M}{k} = M-\left(\left\lceil\frac{k}{r}\right\rceil-1\right)r\frac{M}{k}= M-\left\lceil\frac{k}{r}\right\rceil \frac{M}{k}+r\frac{M}{k}$.
To cover that we need an additional number of blocks
$s \ge
\left\lceil\frac{M-l'r\frac{M}{k}}{\frac{M}{k}}\right\rceil=k-l'r = k-\left(\left\lceil\frac{k}{r}\right\rceil-1\right)r.$
Hence, our final set $\mathcal{S}_{l}$ has size
{\small
\begin{equation}
\begin{split}
&\left|\mathcal{S}_{l}\right|+s-1= l(r+1)+s-1\ge l'(r+1)+k-\left(\left\lceil\frac{k}{r}\right\rceil-1\right)-1 \\
&=\left(\left\lceil\frac{k}{r}\right\rceil-1\right)(r+1)+k-r\left(\left\lceil\frac{k}{r}\right\rceil-1\right)-1 =\left\lceil\frac{k}{r}\right\rceil+k-2. \nonumber
\end{split}
\end{equation}
}Again, due to the fact that the distance is bounded by $n-|\mathcal{S}|$ we have
$d\le n-\left\lceil\frac{k}{r}\right\rceil-k+2$.
\hfill$\Box$

From the above proof we obtain the following corollary.
\begin{cor}
In terms of the code distance, non-overlapping $(r+1)$-groups are optimal.
\end{cor}

In \cite{Yekhanin}, it was proven that $(k,n-k)$ linear codes have minimum code distance that is bounded as
$d\le n-k-\left\lceil\frac{k}{r}\right\rceil+2$.
As we see from our distance-locality bound, the limit of linear codes is information theoretic optimal, i.e., linear codes suffice to achieve it.
Indeed, in the following we show that the distance bound is tight and we present randomized and explicit codes that achieve it.\footnote{In our following achievability proof of the above information theoretic bound we assume that $(r+1)|n$ and we consider non-overlapping repair groups.
This means that $\Gamma(i)\equiv \Gamma(j)$ for all $i,j\in \Gamma(i)$.}

\section{Achievability of the Bound}

In this section, we show that the bound of Theorem \ref{theo:distance} is  achievable using a random linear network coding (RLNC) approach as the one presented in \cite{Ho1}
Our proof uses a variant of the information flow graph that was introduced in \cite{dimakis2010network}. 
We show that a distance $d$ is feasible if a cut-set bound on this new flow graph is sufficiently large for multicast sessions to run on it.

In the same manner as \cite{dimakis2010network}, the information flow graph represents a network where the $k$ input blocks are depicted as sources, the $n$ coded blocks are represented as intermediate nodes of the network, and the sinks of the network are nodes that need to decode the $k$ file blocks.
The innovation of the new  flow graph is that it is ``locality aware'' by incorporating an appropriate dependency subgraph that accounts for the existence of repair groups of size $(r+1)$.
The specifications of this network, i.e., the number and degree of blocks, the edge-capacities, and the cut-set bound are all determined by the code parameters $k,n-k,r,d$. 
For coding parameters that do not violate the distance bound in Theorem \ref{theo:distance}, the
minimum $s-t$ cut of such a flow graph is at least $M$.
The multicast capacity of the induced network is achievable using random linear network codes.
This achievability scheme corresponds to a {\it scalar linear} code with parameters $k,n-k,r,d$. 

\begin{figure}[th]
\centerline{\includegraphics[width=1.05\columnwidth]{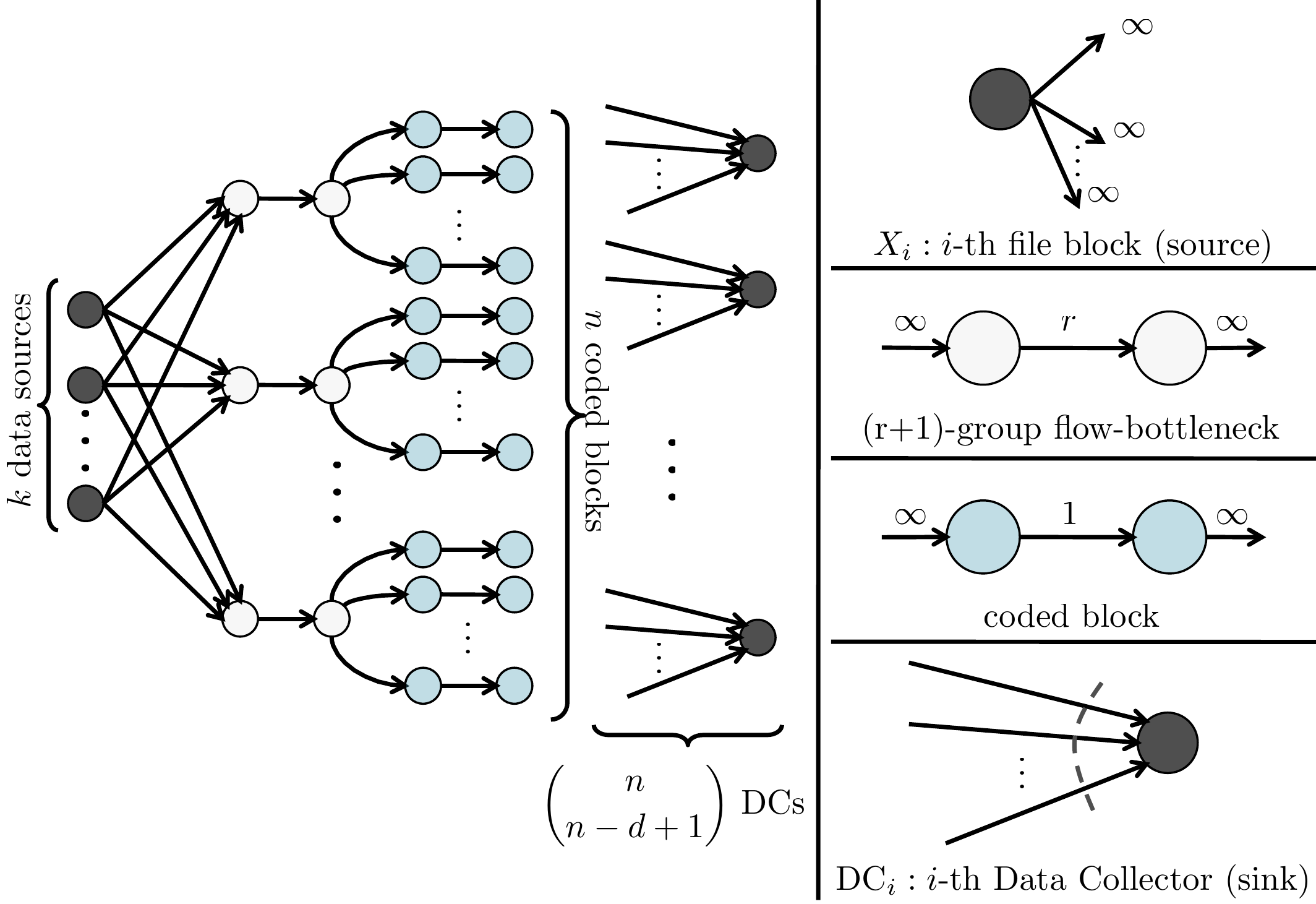}}
\caption{The $\mathcal{G}(k,n-k,r,d)$ information flow graph.}
\label{fig:flow_graph}
\end{figure}

In Fig. \ref{fig:flow_graph}, we show the general structure of an information flow graph.
We refer to this {\it directed} graph as $\mathcal{G}(k,n-k,r,d)$ with vertex set
{
\begin{align}
\mathcal{V} = 
&\left\{\{X_i;i\in[k]\},\,\;\left\{\Gamma_j^{\text{in}},\Gamma_j^{\text{out}}; j\in[n]\right\},\right.\nonumber\\
&\left.\left\{Y_j^{\text{in}},Y_j^{\text{out}}; j\in[n]\right\}, \left\{\text{DC}_l;\forall l\in[T]\right\}
\right\}.\nonumber
\end{align}
}The directed edge set is implied by the following edge capacity function
{\small
\begin{equation}
c_e(v,u)=\left\{
\begin{array}{c@{}l}
\infty ,& (v,u) \in\left(\{X_i;i\in[k]\}, \left\{\Gamma_j^{\text{in}};j\in\left[\frac{n}{r+1}\right] \right\} \right)\\
&\cup\left(\left\{\Gamma_j^{\text{out}};j\in\left[\frac{n}{r+1}\right] \right\}, \left\{Y_j^{\text{in}}; j\in[n]\right\}\right)\\
   & \cup\left(\left\{Y_j^{\text{out}} j\in[n]\right\},\;\{\text{DC}_l;l\in[T]\}\right),\\
\frac{M}{k}, & (v,u)\in\left(\left\{Y_j^{\text{in}} j\in[n]\right\},\;  \left\{Y_j^{\text{out}} j\in[n]\right\}\right),\\
0, &\text{otherwise}.
\end{array}
\right.\nonumber
\end{equation}
}The vertices $\{X_i;i\in[k]\}$ correspond to the $k$ file blocks and $\left\{Y_j^{\text{out}}; j\in[n]\right\}$ correspond to the coded blocks.
The edge capacity between the in- and out- $Y_i$ vertices corresponds to the entropy of a single coded block.
When, $r+1$ blocks are elements of a group, then their ``joint flow,'' or entropy, cannot exceed $r\frac{M}{k}$.
To enforce this entropy constraint, we  bottleneck the in-flow of each group by a node that restricts it to be at most $r\frac{M}{k}$.
For a group $\Gamma(i)$, we add node $\Gamma_i^{\text{in}}$ that receives flow by the sources and
 is connected with an edge of capacity $r\frac{M}{k}$ to a new node $\Gamma_i^{\text{out}}$.
The latter connects to the $r+1$ blocks of the $i$-th group.
The file blocks travel along the edges of this graph towards the sinks, which we call Data Collectors (DCs).
A DC needs to connect to as many coded blocks as such that it can reconstruct the file.
This is equivalent to requiring $s-t$ cuts between the file blocks and the DCs that are at least equal to $M$, i.e., the file size.
We should note that when we are considering a specific group, we know that any block within that group can be repaired from the remaining $r$ blocks.
When a block is lost, the functional dependence among the blocks in an $(r+1)$-group allow a newcomer block to compute a function on the remaining $r$ blocks and reconstruct what was lost.

Observe that if the distance of the code is $d$, then there are $T={{n}\choose{n-d+1}}$ DCs, each with in-degree $n-d+1$, whose incident vertices originate from $n-d+1$ blocks.
The cut-set bound of this network is defined by the set of minimum cuts between the file blocks and each of the DCs.
A source-DC cut in $\mathcal{G}(k,n-k,r,d)$  determines the amount of flow that travels from the file blocks to the DCs.
When $d$ is consistent with the bound of Theorem \ref{theo:distance}, the minimum of all the $s-t$ cuts is at least as much as the file size $M$.
The following lemma states that if $d$ is consistent with the bound of Theorem \ref{theo:distance}, then the minimum of all the cuts is at least as much as the file size $M$.
\begin{lem}
The minimum source-DC cut in $\mathcal{G}(k,n-k,r,d)$ is at least $M$, when $d\le n-\left\lceil\frac{k}{r}\right\rceil-k+2$.
\label{lem:mincut}
\end{lem}
{\bf Proof }: Omitted due to lack of space. \hfill$\Box$\\
Lemma \ref{lem:mincut} verifies that for given $n,k,r$, and a valid distance $d$ according to Theorem \ref{theo:distance}, the information flow graph is consistent with the bound: the DCs have enough entropy to decode all file blocks, when the minimum cut is more than $M$.
The above results imply that the flow graph $\mathcal{G}(k,n-k,r,d)$ captures both the blocks locality and the DC requirements.
Then, a successful multicast session on $\mathcal{G}(k,n-k,r,d)$ is equivalent to all DCs decoding the file.
\begin{theo}
If a multicast session on $\mathcal{G}(k,n-k,r,d)$ is feasible, then there exist a $(k,n-k)$ code $\mathcal{C}$ of locality $r$ and  distance $d$ .
\end{theo}

Hence, the random linear network coding (RLNC) scheme of Ho {\it et al.} \cite{Ho1} achieves the cut-set bound of $\mathcal{G}_r(k,n-k,r,d)$, i.e., there exist capacity achieving network codes, which implies that there exist codes that achieve the distance bound of Theorem \ref{theo:distance}.
Instead of the RLNC scheme, we could use the deterministic construction algorithm of Jaggi {\it et al.} ~\cite{Jaggi} to construct explicit capacity achieving linear codes for multicast networks.
Using that scheme, we could obtain in time polynomial in $T$ explicit $(k,n-k)$ codes of locality $r$.
\begin{lem}
\label{lem:RLNC}
For a network with $E$ edges, $k$ sources, and $T$ destinations, where $\eta$ links transmit linear combination of inputs, the probability of success of the RLNC scheme is at least
$\left(1-\frac{T}{q}\right)^\eta$.
Moreover, using the algorithm in \cite{Jaggi},  a deterministic linear code over $\mathbb{F}$ can be found in time 
$\mathcal{O}\left(ETk(k+T)\right)$.
\end{lem}
The number of edges in our network is 
$E= \frac{n(k+2r+3)}{r+1}+(n-d+1){{n}\choose{k+\left\lceil\frac{k}{r}\right\rceil-1}}$
hence we can calculate the complexity order of the deterministic algorithm, which is
$ETk(k+T) = \mathcal{O}\left(T^3k^2\right)=\mathcal{O}\left(k^28^{nH_2\left(\frac{r}{(r+1)R}\right)}\right)$,
where $H_2(\cdot)$ is the binary entropy function.
The above and Lemma \ref{lem:RLNC} give us the following existence theorem
\begin{theo}
There exists a linear code over $\mathbb{F}$ with locality $r$ and length $n$, such that $(r+1)|n$, that has distance $d=n-\left\lceil\frac{k}{r}\right\rceil-k+2$, if
$|\mathbb{F}|=q>{{n}\choose{k+\left\lceil\frac{k}{r}\right\rceil-1}}=\mathcal{O}\left(2^{nH_2\left(\frac{r}{(r+1)R}\right)}\right).$
Moreover, we can construct explicit codes over $\mathbb{F}$, with $|\mathbb{F}|=q$, in time 
$\mathcal{O}\left(k^28^{nH_2\left(\frac{r}{(r+1)R}\right)}\right)$.
\label{theo:existence}
\end{theo}
Observe that by setting $r=\log(k)$, we obtain Theorem \ref{theo:LRC}.
Moreover, we would like to note that if for each $(r+1)$-group we ``deleted'' a coded block, then the remaining code would be a $(k,n'-k)$-MDS code, where $n'=n-\frac{n}{r+1}$, assuming no repair group overlaps.
This means that LRCs are constructed on top of MDS codes by adding $r$-degree parity coded blocks.
A general construction that operated over small fields and could be constructed in time polynomial in the number of DCs is an interesting open problem.

\section{An Explicit LRC using Reed-\\ Solomon Parities}

We design a $(10,6,5)$-LRC based on Reed-Solomon Codes and Interference Alignment.
We use as a basis for that a $(10,4)$-RS code defined over a binary extension field $\mathbb{F}_{2^m}$.
We concentrate on these specific instances of RS codes since these are the ones that are implemented in practice and in particular in 
the HDFS RAID component of Hadoop.
We continue introducing a general framework for the desing of $(k,n-k)$ Reed-Solomon Codes.

The $k\times n$ (Vandermonde type) parity-check matrix of a $(k,n-k)$-RS code defined over an extended binary field $\mathbb{F}_{2^m}$, of order $q=2^m$, is given by
$[{\bf H}]_{i,j} =a^{i-1}_{j-1}$,
where $a_0,a_1,\ldots,a_{n-1}$ are $n$ distinct elements of the field $\mathbb{F}_{2^m}$.
The order of the field has to be $q\ge n$.
The $n-1$ coefficients $a_0,a_1,\ldots,a_{n-1}$ are $n$ {\it distinct} elements of the field $\mathbb{F}_{2^m}$.
We can select $\alpha$ to be a generator element of the cyclic multiplicative group defined over $\mathbb{F}_{2^m}$.
Hence, let $\alpha$ be a primitive element of the field $\mathbb{F}_{2^m}$.
Then, $[{\bf H}]_{i,j} = \alpha^{(i-1)(j-1)}$, for $i\in[k],j\in[n]$.
The above parity check matrix defines a $(k,n-k)$-RS code.
It is a well-known fact, that due to its determinant structure, any $(n-k)\times(n-k)$ submatrix of ${\bf H}$ has a nonzero determinant, hence, is full-rank.
This, in terms, means that a $(k,n-k)$-RS defined using the parity check matrix ${\bf H}$ is an MDS code, i.e., has optimal minimum distance $d=n-k+1$.
We refer to the $k\times n$ generator matrix of this code as ${\bf G}$.

Based on a $(14,10)$-RS generator matrix, we will introduce $2$ simple parities on the first $5$ and second $5$ coded blocks of the RS code.
This, will yield the generator matrix of our LRC
\begin{equation}
{\bf G}_{\text{LRC}} = \left[{\bf G}\left|\sum_{i=1}^5{\bf g}_i \;\;\sum_{i=6}^{10}{\bf g}_i \right.\right],
\end{equation}
where ${\bf g}_i$ denotes the $i$-th column of ${\bf G}$, for $i\in[14]$.
We would like to note that even if ${\bf G}_{\text{LRC}}$ is not in systematic form, i.e., the first $10$ blocks are not the initial file blocks, 
we can easily convert it into one.
To do so we need to apply a full-rank transformation on the rows of ${\bf G}_{\text{LRC}}$ in the following way:
${\bf A}{\bf G}_{\text{LRC}}
= {\bf A}\left[{\bf G}_{:,1:10} \;\;{\bf G}_{:,11:15}\right]
= \left[{\bf I}_{10} \;\;{\bf A}{\bf G}_{:,11:15}\right]$,
where ${\bf A}={\bf G}_{:,1:10}^{-1}$ and ${\bf G}_{:,i:j}$ is a submatrix of ${\bf G}$ that consists of columns with indices from $i$ to $j$.
This transformation renders our code systematic, while retaining its distance and locality properties.
We proceed to the main result of this section. 


\begin{theo}
The code $\mathcal{C}$ of length 16 defined by ${\bf G}_{\text{LRC}}$ has locality $5$ for all coded blocks and optimal distance $d=5$.
\end{theo}

{\bf Proof:}
We first prove that all coded blocks of ${\bf G}_{\text{LRC}}$ have locality $5$.
Instead of considering block locality, we can equivalently consider the locality of the columns of ${\bf G}_{\text{LRC}}$, without loss of generality.
First let $i\in[5]$.
Then, ${\bf g}_i$ can be reconstructed from the XOR parity $\sum_{j=1}^{5}{\bf g}_j $ if the $4$ other columns ${\bf g}_i$, $j\in\{6,\ldots,10\}\backslash i$, are subtracted from it.
The same goes for $i\in\{6,\ldots, 10\}$, i.e., ${\bf g}_i$ can be reconstructed by subtracting ${\bf g}_j$, for $j\in\{6,\ldots,10\}\backslash i$, from the XOR parity $\sum_{j=6}^{10}{\bf g}_j $.
However, it is not straightforward how to repair the last $4$ coded blocks, i.e., the parity blocks of the systematic code representation.
At this point we make use of Interference Alignment.
Specifically, we observe the following: since the all-ones vector of length $n$ is in the span of the rows of the parity check matrix ${\bf H}$, then it has to be orthogonal to the generator matrix ${\bf G}$, i.e., ${\bf G}{\bf 1}^T={\bf 0}_{k\times 1}$ due to the fundamental property
${\bf G}{\bf H}^T = {\bf 0}_{k\times (n-k)}$.
\balance
This means that 
${\bf G}{\bf 1}^T= {\bf 0}_{k\times 1}\Leftrightarrow \sum_{i=1}^{14}{\bf g}_i = {\bf 0}_{k\times 1}$ and
any columns  of ${\bf G}_{\text{LRC}}$ between the $11$-th and $14$-th are also a function of $5$ other columns.
For example, for $Y_{11}$ observe that we have
${\bf g}_{11}=\left(\sum_{i=1}^{5}{\bf g}_i\right)+\left(\sum_{i=6}^{10}{\bf g}_i\right) +{\bf g}_{12}+{\bf g}_{13}+{\bf g}_{14}$,
where $\left(\sum_{i=1}^{5}{\bf g}_i\right)$ is the first XOR parity and $\left(\sum_{i=6}^{10}{\bf g}_i\right) $ is the second and ``$-$''s become ``$+$''s due to the binary extended field.
In the same manner as ${\bf g}_{11}$, all other columns can be repaired using $5$ columns of ${\bf G}_{\text{LRC}}$.
Hence all coded blocks have locality $5$.

It should be clear that the distance of our code is at least equal to its $(14,10)$-RS precode, that is, $d\ge5$.
We prove that $d=5$ is the maximum distance possible for a length $16$ code has block locality $5$.
Let all codes of locality $r=5$ and length $n=16$ for $M=10$.
Then, there exist $6$-groups associated with the $n$ coded blocks of the code.
Let, $Y_{\Gamma(i)}$ be the set of 6 coded blocks in the repair group of $i \in[16]$.
Then, 
$H(Y_{\Gamma(i)})\le 5$,
for all $i\in[16]$.
Moreover, observe that due to the fact that $5\not|16$ there have to exist at least two distinct overlapping groups $Y_{\Gamma(i_1)}$ and $Y_{\Gamma(i_2)}$, $i_1,i_2\in[16]$, such that
$\left|Y_{\Gamma(i_1)}\cap Y_{\Gamma(i_2)}\right|\ge 1$.
Hence, although the cardinality of $\left|Y_{\Gamma(i_1)}\cup Y_{\Gamma(i_2)}\right|$ is $11$ its joint entropy is bounded as
$H(Y_{\Gamma(i_1)}, Y_{\Gamma(i_2)})= H(Y_{\mathcal{R}(i_1)})+H(Y_{\mathcal{R}(i_2)}|Y_{\mathcal{R}(i_1)})<10$, i.e.,
 at least one additional coded block has to be included to reach an aggregate entropy of $M=10$.
Therefore, any code of length $n=16$ and locality $5$ can have distance at most $5$, i.e.,
$d=5$ is optimal for the given locality.\hfill$\Box$

\end{appendix}


\begin{thebibliography}{10}

\bibitem{amazonec2}
{Amazon EC2}.
\newblock {\em http://aws.amazon.com/ec2/}.

\bibitem{hdfsraidwiki}
{HDFS-RAID} wiki.
\newblock {\em http://wiki.apache.org/hadoop/HDFS-RAID}.

\bibitem{cadambeAsymptotic}
V.~Cadambe, S.~Jafar, H.~Maleki, K.~Ramchandran, and C.~Suh.
\newblock Asymptotic interference alignment for optimal repair of mds codes in
  distributed storage.
\newblock {\em Submitted to IEEE Transactions on Information Theory, Sep. 2011
  (consolidated paper of arXiv:1004.4299 and arXiv:1004.4663)}.

\bibitem{Azure_Storage}
B.~Calder, J.~Wang, A.~Ogus, N.~Nilakantan, A.~Skjolsvold, S.~McKelvie, Y.~Xu,
  S.~Srivastav, J.~Wu, H.~Simitci, et~al.
\newblock Windows azure storage: A highly available cloud storage service with
  strong consistency.
\newblock In {\em Proceedings of the Twenty-Third ACM Symposium on Operating
  Systems Principles}, pages 143--157, 2011.

\bibitem{Orchestra}
M.~Chowdhury, M.~Zaharia, J.~Ma, M.~I. Jordan, and I.~Stoica.
\newblock Managing data transfers in computer clusters with orchestra.
\newblock In {\em SIGCOMM-Computer Communication Review}, pages 98--109, 2011.

\bibitem{dimakis2010network}
A.~Dimakis, P.~Godfrey, Y.~Wu, M.~Wainwright, and K.~Ramchandran.
\newblock Network coding for distributed storage systems.
\newblock {\em IEEE Transactions on Information Theory}, pages 4539--4551,
  2010.

\bibitem{dimakis2011survey}
A.~Dimakis, K.~Ramchandran, Y.~Wu, and C.~Suh.
\newblock A survey on network codes for distributed storage.
\newblock {\em Proceedings of the IEEE}, 99(3):476--489, 2011.

\bibitem{diskreduce}
B.~Fan, W.~Tantisiriroj, L.~Xiao, and G.~Gibson.
\newblock Diskreduce: Raid for data-intensive scalable computing.
\newblock In {\em Proceedings of the 4th Annual Workshop on Petascale Data
  Storage}, pages 6--10. ACM, 2009.

\bibitem{ford2010availability}
D.~Ford, F.~Labelle, F.~Popovici, M.~Stokely, V.~Truong, L.~Barroso, C.~Grimes,
  and S.~Quinlan.
\newblock Availability in globally distributed storage systems.
\newblock In {\em Proceedings of the 9th USENIX conference on Operating systems
  design and implementation}, pages 1--7, 2010.

\bibitem{Yekhanin}
P.~Gopalan, C.~Huang, H.~Simitci, and S.~Yekhanin.
\newblock On the locality of codeword symbols.
\newblock {\em CoRR}, abs/1106.3625, 2011.

\bibitem{greenan2009reliability}
K.~Greenan.
\newblock {\em Reliability and power-efficiency in erasure-coded storage
  systems}.
\newblock PhD thesis, University of California, Santa Cruz, December 2009.

\bibitem{greenan2010mean}
K.~Greenan, J.~Plank, and J.~Wylie.
\newblock {Mean time to meaningless: MTTDL, Markov models, and storage system
  reliability}.
\newblock In {\em HotStorage}, 2010.

\bibitem{Costofcloud}
A.~Greenberg, J.~Hamilton, D.~A. Maltz, and P.~Patel.
\newblock The cost of a cloud: Research problems in data center networks.
\newblock {\em Computer Communications Review (CCR)}, pages 68--73, 2009.

\bibitem{VL2}
A.~Greenberg, J.~R. Hamilton, N.~Jain, S.~Kandula, C.~Kim, P.~Lahiri, D.~A.
  Maltz, P.~Patel, and S.~Sengupta.
\newblock V{L}2: A scalable and flexible data center network.
\newblock {\em SIGCOMM Comput. Commun. Rev.}, 39:51--62, Aug. 2009.

\bibitem{DCell}
C.~Guo, H.~Wu, K.~Tan, L.~Shi, Y.~Zhang, and S.~Lu.
\newblock D{C}ell: a scalable and fault-tolerant network structure for data
  centers.
\newblock {\em SIGCOMM Comput. Commun. Rev.}, 38:75--86, August 2008.

\bibitem{Ho1}
T.~Ho, M.~M\'{e}dard, R.~Koetter, D.~Karger, M.~Effros, J.~Shi, and B.~Leong.
\newblock A random linear network coding approach to multicast.
\newblock {\em IEEE Transactions on Information Theory}, pages 4413--4430,
  October 2006.

\bibitem{Pyramid}
C.~Huang, M.~Chen, and J.~Li.
\newblock Pyramid codes: Flexible schemes to trade space for access efficiency
  in reliable data storage systems.
\newblock {\em NCA}, 2007.

\bibitem{Jaggi}
S.~Jaggi, P.~Sanders, P.~A. Chou, M.~Effros, S.~Egner, K.~Jain, and
  L.~Tolhuizen.
\newblock Polynomial time algorithms for multicast network code construction.
\newblock {\em Information Theory, IEEE Transactions on}, 51(6):1973--1982,
  2005.

\bibitem{Khan_Fast12}
O.~Khan, R.~Burns, J.~Plank, W.~Pierce, and C.~Huang.
\newblock Rethinking erasure codes for cloud file systems: Minimizing {I/O} for
  recovery and degraded reads.
\newblock In {\em FAST 2012}.

\bibitem{kbp:11:iso}
O.~Khan, R.~Burns, J.~S. Plank, and C.~Huang.
\newblock In search of {I/O}-optimal recovery from disk failures.
\newblock In {\em HotStorage '11: 3rd Workshop on Hot Topics in Storage and
  File Systems}, Portland, June 2011. USENIX.

\bibitem{Writeoff}
D.~Narayanan, A.~Donnelly, and A.~Rowstron.
\newblock Write off-loading: Practical power management for enterprise storage.
\newblock {\em ACM Transactions on Storage (TOS)}, 4(3):10, 2008.

\bibitem{Oggier}
F.~Oggier and A.~Datta.
\newblock Self-repairing homomorphic codes for distributed storage systems.
\newblock In {\em INFOCOM, 2011 Proceedings IEEE}, pages 1215 --1223, april
  2011.

\bibitem{DimitrisISIT}
D.~Papailiopoulos and A.~G. Dimakis.
\newblock Locally repairable codes.
\newblock In {\em ISIT 2012}.

\bibitem{papailiopoulos2011simple}
D.~Papailiopoulos, J.~Luo, A.~Dimakis, C.~Huang, and J.~Li.
\newblock Simple regenerating codes: Network coding for cloud storage.
\newblock {\em Arxiv preprint arXiv:1109.0264}, 2011.

\bibitem{Rashmi1}
K.~Rashmi, N.~Shah, and P.~Kumar.
\newblock Optimal exact-regenerating codes for distributed storage at the msr
  and mbr points via a product-matrix construction.
\newblock {\em Information Theory, IEEE Transactions on}, 57(8):5227 --5239,
  aug. 2011.

\bibitem{rashmi2011optimal}
K.~Rashmi, N.~Shah, and P.~Kumar.
\newblock Optimal exact-regenerating codes for distributed storage at the msr
  and mbr points via a product-matrix construction.
\newblock {\em Information Theory, IEEE Transactions on}, 57(8):5227--5239,
  2011.

\bibitem{ReedSolomon}
I.~Reed and G.~Solomon.
\newblock Polynomial codes over certain finite fields.
\newblock In {\em Journal of the SIAM}, 1960.

\bibitem{rodrigues2005high}
R.~Rodrigues and B.~Liskov.
\newblock High availability in dhts: Erasure coding vs. replication.
\newblock {\em Peer-to-Peer Systems IV}, pages 226--239, 2005.

\bibitem{shah2012interference}
N.~Shah, K.~Rashmi, P.~Kumar, and K.~Ramchandran.
\newblock Interference alignment in regenerating codes for distributed storage:
  Necessity and code constructions.
\newblock {\em Information Theory, IEEE Transactions on}, 58(4):2134--2158,
  2012.

\bibitem{Tamo}
I.~Tamo, Z.~Wang, and J.~Bruck.
\newblock {MDS} array codes with optimal rebuilding.
\newblock {\em CoRR}, abs/1103.3737, 2011.

\bibitem{Wicker}
S.~B. Wicker and V.~K. Bhargava.
\newblock Reed-solomon codes and their applications.
\newblock In {\em IEEE Press}, 1994.

\bibitem{xin2003reliability}
Q.~Xin, E.~Miller, T.~Schwarz, D.~Long, S.~Brandt, and W.~Litwin.
\newblock Reliability mechanisms for very large storage systems.
\newblock In {\em MSST}, pages 146--156. IEEE, 2003.

\end{thebibliography}
\end{document}